\documentclass[fleqn,usenatbib]{mnras}

\usepackage{newtxtext,newtxmath}

\usepackage[T1]{fontenc}

\DeclareRobustCommand{\VAN}[3]{#2}
\let\VANthebibliography\thebibliography
\def\thebibliography{\DeclareRobustCommand{\VAN}[3]{##3}\VANthebibliography}

\usepackage{graphicx}	
\usepackage{amsmath}	
\usepackage{orcidlink}  
\usepackage{hyperref}   
\usepackage{caption}    
\usepackage{subcaption} 
\usepackage{CJKutf8}

\title[Resolved SF Activity with MAGPI]{The MAGPI Survey: Evolution of radial trends in star formation activity across cosmic time}

\author[Marcie Mun et al.]{
Marcie Mun,$^{\orcidlink{0000-0002-3706-9955} 1,2}$\thanks{E-mail: jaeyeon.mun@anu.edu.au}
Emily Wisnioski,$^{\orcidlink{0000-0003-1657-7878} 1,2}$
Andrew J. Battisti,$^{\orcidlink{0000-0003-4569-2285} 1,2}$
J. Trevor Mendel,$^{\orcidlink{0000-0002-6327-9147} 1,2}$
Sara L. Ellison,$^{\orcidlink{0000-0002-1768-1899} 3}$
\newauthor Edward N. Taylor,$^{4}$
Claudia D. P. Lagos,$^{\orcidlink{0000-0003-3021-8564} 2,5}$
Katherine E. Harborne,$^{\orcidlink{0000-0002-2043-7985} 2,5}$
Caroline Foster,$^{\orcidlink{0000-0003-0247-1204} 6,2}$
\newauthor Scott M. Croom,$^{\orcidlink{0000-0003-2880-9197} 7,2}$
Sabine Bellstedt,$^{\orcidlink{0000-0003-4169-9738} 5}$
Stefania Barsanti,$^{\orcidlink{0000-0002-9332-5386} 1,2,7}$
Anshu Gupta,$^{\orcidlink{0000-0002-8984-3666} 8,2}$
Lucas M. Valenzuela,$^{\orcidlink{0000-0002-7972-9675} 9}$
\newauthor Qian-Hui Chen (陈千惠),$^{\orcidlink{0000-0002-4382-1090} 1,2}$
Kathryn Grasha,$^{\orcidlink{0000-0002-3247-5321} 1,2}$
Tamal Mukherjee,$^{\orcidlink{0009-0004-7639-869X} 10,2}$
Hye-Jin Park,$^{\orcidlink{0000-0002-9809-6631} 1,2}$
Piyush Sharda,$^{\orcidlink{0000-0003-3347-7094} 11}$
\newauthor Sarah M. Sweet,$^{\orcidlink{0000-0002-1576-2505} 12,2}$
Rhea-Silvia Remus,$^{9}$
Tayyaba Zafar$^{\orcidlink{0000-0003-3935-7018} 10,13}$
\\
$^{1}$Research School of Astronomy and Astrophysics, Australian National University, Weston Creek, ACT 2611, Australia\\
$^{2}$ARC Centre of Excellence for All Sky Astrophysics in 3 Dimensions (ASTRO 3D), Canberra, ACT 2611, Australia\\
$^{3}$Department of Physics and Astronomy, University of Victoria, Finnerty Road, Victoria, British Columbia, V8P 1A1, Canada\\
$^{4}$Centre for Astrophysics and Supercomputing, Swinburne University of Technology, John Street, Hawthorn, 3122, Australia\\
$^{5}$International Centre for Radio Astronomy Research, The University of Western Australia, 35 Stirling Highway, Crawley, WA 6009, Australia\\
$^{6}$School of Physics, University of New South Wales, Sydney, NSW 2052, Australia\\
$^{7}$Sydney Institute for Astronomy, School of Physics, University of Sydney, NSW 2006, Australia\\
$^{8}$International Centre for Radio Astronomy Research (ICRAR), Curtin University, Bentley WA, Australia\\
$^{9}$Universitäts-Sternwarte München, Fakultät für Physik, Ludwig-Maximilians-Universität München, Scheinerstr. 1, 81679 München, Germany\\
$^{10}$School of Mathematical and Physical Sciences, Macquarie University, NSW 2109, Australia\\
$^{11}$Leiden Observatory, Leiden University, P.O. Box 9513, NL-2300 RA Leiden, The Netherlands\\
$^{12}$School of Mathematics and Physics, University of Queensland, Brisbane, QLD 4072, Australia\\
$^{13}$Macquarie University Astrophysics and Space Technologies Research Centre, Sydney, NSW 2109, Australia
}

\date{Accepted XXX. Received YYY; in original form ZZZ}

\pubyear{2023}

\begin{document}
\label{firstpage}
\begin{CJK}{UTF8}{gbsn}
\pagerange{\pageref{firstpage}--\pageref{lastpage}}
\maketitle

\begin{abstract}
Using adaptive optics with the Multi-Unit Spectroscopic Explorer (MUSE) on the Very Large Telescope (VLT), the Middle Ages Galaxy Properties with Integral Field Spectroscopy (MAGPI) survey allows us to study the spatially resolved Universe at a crucial time of $\sim$4 Gyr ago ($z \sim$ 0.3) when simulations predict the greatest diversity in evolutionary pathways for galaxies. We investigate the radial trends in the star formation (SF) activity and luminosity-weighted stellar ages as a function of offset from the star-forming main sequence (SFMS) for a total of 294 galaxies. Using both H$\alpha$ emission and the 4000 \AA \ break (i.e., D4000) as star formation rate (SFR) tracers, we find overall flat radial profiles for galaxies lying on and above the SFMS, suggestive of physical processes that enhance/regulate SF throughout the entire galaxy disc. However, for galaxies lying below the SFMS, we find positive gradients in SF suggestive of inside-out quenching. Placing our results in context with results from other redshift regimes suggests an evolution in radial trends at $z \sim$ 0.3 for SF galaxies above the SFMS, from uniformly enhanced SF at $z\sim1$ and $z \sim$ 0.3 to centrally enhanced SF at $z \sim$ 0 (when averaged over a wide range of mass). We also capture higher local SFRs for galaxies below the SFMS compared to that of $z \sim$ 0, which can be explained by a larger population of quenched satellites in the local Universe and/or different treatments of limitations set by the D4000-sSFR relation. 
\end{abstract}




\section{Introduction}
\label{sec:intro}
Galaxies are some of the most complex structures in the Universe. The formation and evolutionary pathways galaxies take throughout their lifetimes are multifold. Despite the diversity of such pathways, populations of galaxies as a whole have been observed to follow tight scaling relations across cosmic time, such as the correlation between star formation rates (SFRs) and stellar masses (M\textsubscript{$\star$}), otherwise known as the star-forming main sequence (SFMS; e.g., \citealt{elbaz2007, noeske2007, whitaker2012, lee2015, renzinipeng2015}). The existence of such a tight relation ($\sim$0.2 -- 0.35 dex in scatter) observed up to redshifts as high as $z \sim$ 6 \citep[e.g.,][]{speagle2014, thorne2021, popesso2023} suggests that galaxies are able to maintain their star formation in a quasi-steady manner across cosmic time \citep[e.g.,][]{bouche2010}. Galaxies not on the SFMS are considered to be outliers, where galaxies above the main sequence are undergoing starbursts and galaxies below are either in the process of or have already quenched their star formation. While the SFMS has been in place across cosmic time, the normalisation of the relation has evolved together with the cosmic star formation rate density \citep{madaudickinson2014, thorne2021, popesso2023, dsilva2023}. This suggests that there may be differences in the star-forming conditions between the early Universe and today, such that galaxies hosted much more gas-rich and star-forming discs in the past \citep[e.g.,][]{schminovich2005, daddi2010, tacconi2010, decarli2016, decarli2019}. Scaling relations between various galaxy properties, along with multi-wavelength studies of galaxies at different redshift regimes, provide us with the big picture of galaxy evolution across cosmic time. However, in order to understand the processes governing these global relationships, we need to spatially resolve galaxies and their environments to uncover their underlying physics.

Various mechanisms are known to regulate the star formation activity of galaxies, which include those pertaining to internal structures such as bars (e.g., \citealt{ellison2011, cheung2013, vera2016, kim2017, frasermckelvie2020, lin2020}), spiral arms \citep[e.g.,][]{yu2021}, and bulges (i.e., `morphological quenching'; \citealt{martig2009, mendezabreu2019, gensior2020}), active galactic nuclei (AGN; e.g., \citealt{bower2006, croton2006, lagos2008}; see \citealt{fabian2012} for a review), and environmental processes such as interactions with neighbouring galaxies \citep{heckman1990, moore1996, patton2013, davies2015, moreno2021}, some of which can lead to mergers \citep[e.g.][]{hopkins2013, hani2020, ellison2022}, gas stripping due to the surrounding medium \citep{gunngott1972, nulsen1982, kk2004a, kk2004b, bg2006}, and starvation/strangulation \citep{larson1980, bekki2002}. By nature, bars and AGN are expected to have a more noticeable impact on the galaxy centre, either potentially leading to funnelling of gas towards the centre and consequently a nuclear starburst \citep{shlosman1989, emsellem2015}, or the outflow of gas followed by earlier onset of quenching \citep{dimatteo2005, feruglio2010}. AGN can also suppress star formation on a global scale by injecting kinetic energy into the gas distributed throughout the halo, which leads to both a decrease in cold gas inflow and increase in stability of the gas against fragmentation \citep{db2006, dekel2009}. On the other hand, environmental processes add more complexities to the overall picture, where ram pressure stripping acts to preferentially strip gas from galaxy outskirts, quenching the galaxy outside-in \citep{gunngott1972, steinhauser2016}, whereas starvation/strangulation can quench galaxies on a global scale by cutting off cold gas accretion entirely \citep{larson1980, bekki2002, bg2006}. As such, the distribution of the gas, stars, and metals are expected to exhibit variations from the centre towards the outskirts of galaxies \citep{kewley2010, lian2019, sanchez2020}. This is further complicated by the evolution of environments and build-up of the densest environments at late times \citep{knobel2009}.

Distinguishing the different physical mechanisms at play requires a wealth of observational information, among which spatially resolved photometry and/or spectroscopy covering a broad range of wavelengths prove to be crucial. Integral field spectroscopy (IFS) has been particularly useful in this sense, where large integral field unit (IFU) galaxy surveys have used spatially resolved spectroscopy to measure radial variations in galaxy properties. In particular, many studies have analysed radial trends in SFR surface density ($\mathrm{\Sigma_{SFR}}$) and specific star formation rates (sSFR = SFR/M\textsubscript{$\star$}) as a function of morphology, environment, stellar mass, and offset with respect to the SFMS \citep{gonzalezdelgado2016, nelson2016, ellison2018, medling2018, wang2019, bluck2020b}. Such analyses have demonstrated the power of combining global and local information, where comparing radial trends in local star formation as a function of location with respect to the global SFMS has been particularly insightful. 

Studies based on the local Universe ($z \sim 0$) suggests that galaxies above and below the SFMS show enhanced and quenched star formation activity, respectively, in the galaxy centre \citep{ellison2018, wang2019}. Multiple scenarios have been proposed to explain this; for example, a central starburst may precede inside-out quenching in galaxies \citep{ellison2018} due to processes such as galaxy mergers and AGN feedback. On the other hand, it may not necessarily be physical processes, but moreso radial variations in SF efficiencies manifesting in positive/negative gradients in SF \citep{wang2019}. Some studies have attempted to distinguish between internal and external mechanisms -- \citet{bluck2020b} find that central and high-mass satellite galaxies quench inside-out primarily due to AGN feedback, whereas low-mass satellites quench outside-in due to environmental mechanisms such as ram pressure stripping and galaxy-galaxy interactions. Putting radial trends in SF in context with environments ranging from low-mass galaxy groups to clusters have also shown that galaxies show stronger trends of outside-in quenching with increasing environmental density \citep{schaefer2017, schaefer2019, wang2022}. However, features of outside-in quenching (i.e., a negative gradient in SF) can also be interpreted as central rejuvenation due to increased gas accretion rates from galaxy-galaxy interactions, such that both processes might be occurring in concert \citep[e.g.,][]{schaefer2017}. The combination of analysing multiple parameters, such as $\mathrm{\Sigma_{SFR}}$s and sSFRs, has proved helpful towards favouring one scenario over the other \citep[e.g.][]{medling2018}. 

At higher redshifts ($z \sim 1 - 2$), high spatial resolution imaging taken with the \textit{Hubble Space Telescope} (\textit{HST})/Wide Field Camera 3 (WFC3) grism from the 3D-\textit{HST} \citep{skelton2014} survey has made similar analyses possible. Stacked images of both H$\alpha$ and stellar continuum emission show that galaxies above the SFMS are enhanced everywhere throughout the disc, galaxies on the SFMS have on average flat profiles, whereas those below are quenched everywhere \citep{nelson2016}. However, centrally suppressed sSFRs are observed in the centre of their most massive (10.5 $<$ $\mathrm{\log_{10}}$(M\textsubscript{$\star$}/M\textsubscript{\sun}) $<$ 11) galaxies, which suggests signs of inside-out quenching. Similar results are found from the SINS/zC-SINF \citep{forsterschreiber2018} survey with evidence that the most massive ($\mathrm{\log_{10}}$(M\textsubscript{$\star$}/M\textsubscript{\sun}) $\gtrsim$ 11) galaxies are undergoing inside-out quenching, whereas lower mass galaxies show flat sSFR profiles \citep{tacchella2015, tacchella2018}. 

Together, the results at $z = $ 0 and $z = $ 2 suggest an evolution in radial trends over the last $\sim10$ Gyr for galaxies above and below the SFMS. Understanding which changes (e.g. environment, accretion rates, gas fractions, etc.) have contributed to this evolution requires investigating galaxies at the epoch where the cosmic SFRD is winding down. However, studies tracing this evolution have been limited. Observationally, they require large areas and deep high-resolution imaging or imaging spectroscopy. The Multi-Unit Spectroscopic Explorer (MUSE) instrument on the Very Large Telescope (VLT) however is opening up this space with surveys that tile over larger regions of the sky. The natural seeing MUSE-WIDE survey \citep{herenz2017} has made it possible to measure the radial variations in star formation and gas-phase metallicities at 0.1 $< z <$ 0.42, but has been limited in sample statistics \citep{jafariyazani2019,yao2022}. The Middle Ages Galaxy Properties with Integral Field Spectroscopy (MAGPI\footnote{Based on observations obtained at the VLT of the European Southern Observatory (ESO), Paranal, Chile (ESO program ID 1104.B-0526)}; \citealt{foster2021}) survey has been designed to resolve these issues, by achieving comparable spatial resolution to that achieved with local IFS surveys in both stars and ionised gas for $\sim$400 galaxies at $z \sim$ 0.3, with the use of adaptive optics (AO) capabilities on MUSE. 

In this paper, we study the radial trends in the star formation activity of MAGPI galaxies at $z \sim$ 0.3 to address two key questions: (1) Do we observe an evolution in star formation activity with redshift at z $\sim$ 0.3?; and (2) What regulates and/or quenches star formation at $z \sim$ 0.3? We use both the global and resolved SFMS (rSFMS; i.e., resolved counterpart of the global SFMS measured on kiloparsec scales, showing a tight correlation between SFR and stellar mass surface densities; e.g., \citealt{sanchez2013, wuyts2013, canodiaz2016}) to measure the radial variation in local SF. Specifically, we measure radial trends in $\mathrm{\Delta \Sigma_{SFR}}$, the logarithmic offset from the rSFMS, as a function of varying offsets from the global SFMS. We also complement our results with luminosity-weighted stellar age (Age\textsubscript{L}) profiles. We also ensure that our sample is not biased towards star-forming galaxies with strong emission lines by using D4000 to indirectly measure SFRs for passive spaxels/galaxies. In this paper, we introduce the MAGPI survey in more depth in Section \ref{sec:data}, followed by a description of our methodology behind the SFRs and stellar masses, along with global and local star formation metrics in Section \ref{sec:methods}. We show our resulting $\mathrm{\Delta \Sigma_{SFR}}$ and luminosity-weighted age profiles in Section \ref{sec:results}. We then discuss the potential physical mechanisms governing MAGPI galaxies, along with the implications of our results in context with similar studies at different redshift regimes in Section \ref{sec:disc}. Finally, we summarise our results and discuss planned future work in Section \ref{sec:summ}. Throughout this paper, we assume a \citet{chabrier2003} initial mass function (IMF) and adopt a flat $\Lambda$CDM cosmology with $\mathrm{H_{0}}$ = 70 km s\textsuperscript{-1} Mpc\textsuperscript{-1}, $\mathrm{\Omega_{m}}$ = 0.3, and $\mathrm{\Omega_{\Lambda}}$ = 0.7.  

\section{The MAGPI survey}
\label{sec:data}
MAGPI is a VLT/MUSE Large Program (Program ID: 1104.B-0536) aimed at studying the spatially resolved spectroscopy of stars and ionized gas in galaxies at 0.25 $< z <$ 0.35, translating to lookback times of 3 - 4 Gyrs. A main motivation of the MAGPI survey is to trace the morpho-kinematic evolution of galaxies between $z\sim1$ and $z\sim2$. With the aid of Ground Layer Adaptive Optics (GLAO) on MUSE, MAGPI is able to probe both stellar and gas kinematics at a comparable relative spatial resolution to those probed by IFS surveys at lower redshifts, such as MaNGA \citep{bundy2015} and SAMI \citep{croom2012}. The survey targets a total of 60 primary galaxies (M\textsubscript{$\star$} $\mathrm{> 7 \times 10^{10} M_{\sun}}$) and their neighbouring galaxies (`secondary' galaxies; about $\sim$400) within the MUSE field-of-view (FOV) at $z \sim$ 0.3 ($\sim$270 kpc) in a broad range of environments, spanning 12 $\mathrm{\lesssim \log (M_{halo} / M_{\sun}) \lesssim}$ 15 \citep{foster2021}. 

56 of the 60 primary galaxies were sampled from the G12, G15, and G23 fields of the Galaxy and Mass Assembly survey (GAMA; \citealt{driver2011, driver2022}), with the remaining 4 from MUSE archival fields of Abell 370 ($z = $ 0.375; Program ID: 096.A-0710, PI: Bauer) and Abell 2744 ($z = $ 0.308; Program IDs: 095.A-0181 and 096.A-0496, PI: Richard). The sample selection based on the GAMA survey is also supported by the availability of ancillary imaging data ranging from the ultraviolet (UV) to near-infrared (NIR) from the Hyper Suprime Cam Subaru Strategic Program (HSC SSP; \citealt{aihara2019}), Kilo-Degree Survey (KiDS; \citealt{kuijken2019}), and the VISTA Kilo degree Infrared Galaxy (VIKING; \citealt{wright2019}). Centred on a primary target, each of the MAGPI fields covers roughly a 1 $\times$ 1 arcmin FOV with 0.2 arcsecond per pixel spatial sampling. Each field is observed in six observing blocks of 2 $\times$ 1320~s exposures, totaling to 4.4 hours in the wide-field mode (WFM) with GLAO (total of 246 hours on-source). The nominal mode is used such that wavelength coverage ranges from 4700 to 9300 \AA, where spectra are sampled in 1.25~\AA \ pixel\textsuperscript{-1}. The use of GLAO results in a laser gap at 5780 - 6050~\AA \ due to the GALACSI system sodium laser notch filter \citep{hartke2020}. 

Data reduction on the MUSE data are done using \textsc{pymusepipe}\footnote{\url{https://github.com/emsellem/pymusepipe}}, a Python wrapper for the MUSE reduction pipeline \citep{weilbacher2020}. \textsc{pymusepipe} takes care of the major data processing which includes wavelength calibration and measurement of the instrumental line-spread function. \textsc{cubefix} \citep{cantalupo2019} and the Zurich Atmosphere Purge (ZAP; \citealt{soto2016}) are used for illumination correction and sky subtraction, respectively. The post-processed data cubes are then run through \textsc{ProFound}\footnote{\url{https://github.com/asgr/ProFound}} \citep{robotham2018} for source detection, which additionally produces segmentation maps and measures structural parameters such as effective radii and position angles. The `dilated' segmentation maps are used to set the size of data cubes for individual galaxies (i.e., `minicubes') with \textsc{mpdaf}\footnote{\url{https://github.com/musevlt/mpdaf}}, such that the data cubes fully encompass both the full light distribution from the galaxies and sufficient coverage of the background. Dilated segmentation maps are created by expanding the segmentation maps to reach a particular source magnitude limit below the estimated sky level. Further details are provided in \citet{foster2021} and Mendel et al (in prep). 

At the time of writing, observations are still ongoing and a total of 48 fields ($\sim$86 per cent completion rate) have been observed, among which 35 of them have been fully reduced with relevant data products. The 4 archival fields of Abell 370 and Abell 2744 are not included in this analysis. While the main redshift range of interest for MAGPI is centred around $z \sim$ 0.3, for the purposes of sampling as many star-forming galaxies as possible, the redshift range is extended such that the galaxies selected for this study are located at 0.25 $\leq z \leq$ 0.424, where the upper bound for redshift is set by the MUSE detection limit for the H$\alpha$ line. 

\section{Methods}
\label{sec:methods}

\subsection{Emission line fits}
\label{sec:elinefit}
Emission lines are fit using the Galaxy IFU Spectroscopy Tool (\textsc{gist}; \citealt{bittner2019}) code, which is a Python wrapper for \textsc{pPXF} \citep{capem2004, cappellari2017} and \textsc{GandALF} \citep{sarzi2006, falcon-barroso2006}, for which the former is specialized for stellar continuum fitting and the latter for emission line fitting. For stellar continuum measurements, we used the luminosity-weighted spectral templates from the C3K library (Charlie Conroy, private communication) that are based on the Modules for Experiments in Stellar Astrophysics (\textsc{mesa}) Isochrones and Stellar Tracks (\textsc{mist}; \citealt{choi2016}). Prior to fitting, we convolve these templates to match the wavelength-dependent MUSE spectral resolution derived from sky lines (e.g., \citealt{bacon2017}). We have adopted a multiplicative Legendre polynomial of order 12 when fitting the stellar continuum. We adopted a S/N threshold of 10 when Voronoi binning \citep{capcop2003} the stellar continuum; for galaxies with low continuum S/N, we use a single integrated bin. When choosing which spaxels to include in the continuum binning process, we take into account two criteria: 
\begin{enumerate}
    \item Reside inside the dilated \textsc{ProFound} mask
    \item Have integrated S/N $\geq$ 2 for 6050 \AA \ $\mathrm{< \lambda_{obs} <}$ 7750 \AA; corresponding roughly to the peak MUSE sensitivity range
\end{enumerate}
For emission line fitting, we have used a customized version of \textsc{pyGandALF} where errors on fluxes, velocities, and widths for the emission lines are estimated based on a Monte Carlo approach. We also implemented a grid search of initial fitting parameters to improve fitting in the low S/N galaxy outskirts. Velocities of the emission lines are restricted to be within $\pm$600 km/s relative to the stellar continuum velocity of the nearest Voronoi bin and the widths are restricted to be in the range from 1 - 300 km/s. These data products will be made publicly available in the future (Battisti et al. in prep). 

\subsection{Stellar masses}
\label{sec:mstar}
We derive resolved stellar mass maps directly from the MAGPI data itself, which provides uniform coverage across all of the target fields (i.e., GAMA G12, G15, and G23). This is done in two steps. We first derive stellar mass-to-light ratios (M/L) based on spectral energy distribution (SED) fitting to pixel-matched imaging from HSC where available\footnote{The deepest imaging available for MAGPI come from HSC ($\sim$2.1 mag deeper than KiDS in the overlapping \textit{gri} bands on average, in terms of 5$\sigma$ detection limits measured in apertures of 2 arcsec diameters), however HSC imaging is not available in 12/56 of the MAGPI fields.}, which provides broadband coverage over the \textit{grizy} bands following the method described by \citet{taylor2011} and \citet{medling2018}; and summarised briefly below. Next, we construct a relationship between $g-r$ colour and M/L ratio in the $r$-band that can be applied consistently across the full dataset. 

Prior to SED fitting, we match the available imaging in both World Coordinate System (WCS) and pixel scale to the MUSE data cubes. We use the \citet{bc2003} template library to create composite stellar populations (CSPs), where we make assumptions of exponentially declining star formation histories (SFHs) and a single-screen dust attenuation law using \citet{calzetti2000}. A fully Bayesian approach is adopted such that probability-weighted averages are taken as the most likely values of stellar mass per pixel, in contrast to adopting the best-fit value derived via maximum likelihood (see Section 3.4 in \citealt{taylor2011}). We assume uniform priors on a total of 4 parameters - age, e-folding time, metallicity, and dust obscuration. 

The resulting M/L maps are then used to construct a relationship between $g - r$ colour and M/L\textsubscript{\textit{r}}. We do this non-parametrically by taking the median M/L in bins of $g - r$ colour 0.03 mag wide. We adopt the standard deviation of M/L within these bins as indicative of the uncertainty in mapping a single optical colour to M/L, which is of order 0.15 dex regardless of colour. We include this uncertainty in quadrature in our final mass estimates. The colour-M/L relationship derived in this way is applied to rest-frame $g - r$ colour maps computed directly from the MAGPI data themselves using the measured redshifts, and is applied to spaxels that fall within the dilated segmentation maps produced by \textsc{ProFound}. Uncertainties on the MUSE-based rest-frame colours are computed using the uncertainties on the spectral data. The stellar mass maps are derived based on the colour-M/L relationship, where we also make uncertainty cuts on a pixel-by-pixel basis. To choose a reasonable cut, we reference the corresponding S/N resulting from reliable SED fitting of HSC imaging (i.e., S/N $\geq$ 10), which translates roughly into a maximum uncertainty of 0.5 dex. We also measure a total stellar mass by summing the stellar masses per pixel for the pixels that satisfy both the uncertainty and dilated mask cuts. We calculate stellar mass surface densities ($\mathrm{\Sigma_{\star}}$) using: 
\begin{equation} \label{eq:1}
    \mathrm{\Sigma_{\star} (M_{\sun} \ kpc^{-2})} = \frac{\mathrm{M_{\star} (M_{\sun})}}{(0.2 \times \mathrm{D_{A}})^{2}},
\end{equation}
where the MUSE WFM pixel scale is 0.2 arcsec/pixel and $\mathrm{D_{A}}$ refers to the angular diameter distance in kpc/arcsec based on our spectroscopic redshifts and the adopted cosmology.

\subsection{Star formation rates}
\label{sec:sfr} 
The main goal of this study is to understand the different physical mechanisms governing the star formation activity of galaxies in different global star-forming states. It is crucial, therefore, to have a complete sample of measured SFRs such that the sample is not biased towards star-forming galaxies. As such, we use two kinds of SFR indicators for this work - we use the H$\alpha$ line for spaxels/galaxies with high S/N in their emission lines (see Section \ref{sec:ha}) and D4000 for all other spaxels/galaxies (see Section \ref{sec:d4}). Details regarding the methodology for measuring each SFR indicator are given in the following sections. 

\subsubsection{Star-forming spaxels/galaxies}
\label{sec:ha}
\begin{figure}
    \centering
    \includegraphics[width=\columnwidth]{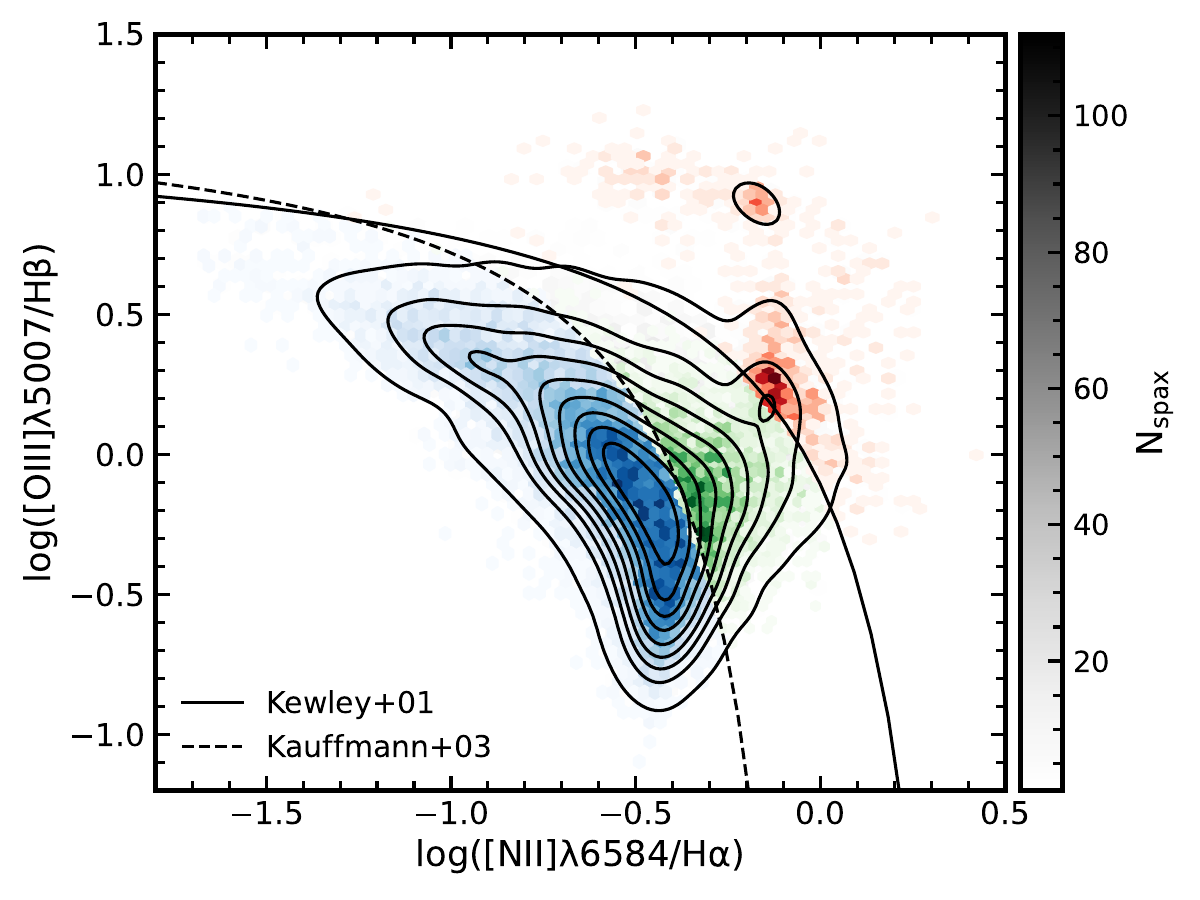}
    \caption{BPT diagram for a total of 16,511 spaxels that satisfy either one of the following conditions: (1) S/N $\geq$ 3 in all 4 emission lines - [OIII]$\lambda$5007, H$\beta$, [NII]$\lambda$6584, and H$\alpha$, or (2) S/N $\geq$ 3 in all except for [NII] and falls in the SF region. 3,055 ($\sim$24 per cent) of the spaxels fall under the second category. Blue denotes the star-forming spaxels, green the composite, and red the AGN. The theoretical and empirical lines from \citet{kewley2001} (solid) and \citet{kauffmann2003} (dashed), respectively, are used to classify spaxels into different regions on the BPT diagram.}
    \label{fig:bpt}
\end{figure}

For galaxies with well-detected emission lines, we use the H$\alpha$ line to estimate SFRs. H$\alpha$ emission traces hydrogen ionized by a variety of sources including star formation, active galactic nuclei (AGN), and shocks. Ionized hydrogen traced by stars roughly translates to star formation timescales of $\sim$10 Myr (e.g., \citealt{ke2012}), i.e., the typical lifetimes of the young and massive O- and B-type stars. As such, H$\alpha$ emission is a recent star formation tracer. In order to ensure that the H$\alpha$ emission is devoid of contamination from other ionization sources (AGN in particular), we use the Baldwin, Phillips \& Terlevich (BPT; \citealt{bpt1981}) emission line diagnostic diagram to identify purely star-forming spaxels within galaxies. For the BPT diagram, shown in Figure~\ref{fig:bpt}, we use the line ratios [OIII]$\lambda$5007/H$\beta$ and [NII]$\lambda$6584/H$\alpha$, and require all 4 emission lines to have S/N $\geq$ 3 for a spaxel to be used in the analysis. However, we observe spaxels with good detection (i.e., S/N $\geq$ 3) in H$\alpha$ but weak (i.e., S/N $<$ 3) signal in [NII]$\lambda$6584, which suggests that we may be missing out on additional SF. To recover as many star-forming spaxels as possible, we take 3 times the error as an upper limit of the [NII]$\lambda$6584 flux \citep[e.g.,][]{rosario2016} for spaxels with S/N $<$ 3 in [NII]$\lambda$6584. We then plot the spaxels with upper limits on the BPT diagram, where we include those that fall in the SF region, whereas those that fall in the composite or AGN regions are excluded entirely from the analysis due to concerns of noise contamination (e.g., this method often results in spaxels in the outskirts being classified as AGN). This results in a total of 16,511 spaxels, where 12,882 (3,055 being upper limits) of them are classified as star-forming. We use the H$\alpha$ emission line to measure SFRs for the 12,882 spaxels classified as star-forming. 

We correct H$\alpha$ flux maps for dust extinction with the Balmer decrement, i.e., the H$\alpha$/H$\beta$ flux ratio. We first correct for foreground extinction due to the Milky Way using the $E(B - V)$ values from \cite{sf2011}. To correct for intrinsic extinction, we adopt the Galactic extinction curve from \citet{fitzpatrick2019} with $R_{V}$ = 3.1 and assume an intrinsic H$\alpha$/H$\beta$ value of 2.86, which corresponds to Case B recombination under the conditions of temperature T = 10\textsuperscript{4} K and electron density $\mathrm{n_{e}}$ = 10\textsuperscript{2} cm\textsuperscript{-3} \citep{osterbrock1989, of2006}. 

The dust-corrected H$\alpha$ fluxes are then converted to luminosities with luminosity distances from measured spectroscopic redshifts. We use the SFR conversion recipe from \citet{kennicutt1998}, which is given below:
\begin{equation} \label{eq:2}
    \mathrm{SFR_{H\alpha}} \ (\mathrm{M_{\sun}} \ \mathrm{yr^{-1}}) = \frac{\mathrm{L_{H\alpha}}}{1.26 \times 10^{41} \ \mathrm{erg \ s^{-1}} \times 1.53},
\end{equation}
where $\mathrm{L_{H\alpha}}$ denotes the H$\alpha$ luminosity. The factor of 1.53 (see Table 1 in \citealt{driver2013}) converts the recipe from a \citet{salpeter1955} to a \citet{chabrier2003} IMF.

We measure local SFR surface densities, $\mathrm{\Sigma_{SFR}}$, for each of the spaxels classified as SF with the following equation:
\begin{equation} \label{eq:3}
    \mathrm{\Sigma_{SFR, H\alpha} (M_{\odot} \ yr^{-1} \ kpc^{-2})} = \frac{\mathrm{SFR_{H\alpha} \ (M_{\sun} \ yr^{-1})}}{(0.2 \times \mathrm{D_{A}})^{2}},
\end{equation}
where the MUSE WFM pixel scale is 0.2 arcsec/pixel and $\mathrm{D_{A}}$ refers to the angular diameter distance in kpc/arcsec. In addition to the spaxel SFR surface densities, we also compute total SFRs for each galaxy by summing up all star-forming spaxels with measured SFRs. Uncertainties on both SFR and $\mathrm{\Sigma_{SFR}}$ are measured via error propagation on the measured uncertainties in H$\alpha$ and H$\beta$ fluxes.

\subsubsection{Passive spaxels/galaxies}
\label{sec:d4}
For galaxies with either weak emission lines or lack of star formation as classified by the BPT diagram, we use D4000 to measure SFRs. Previous studies \citep{brinchmann2004, bluck2020a, bluck2020b, thorp2022} have derived an empirical relation between D4000 and sSFR, that can be used in galaxies with regions either (1) classified as either composite or AGN by the BPT diagram, or (2) with weak S/N in their emission lines. We follow a similar methodology with the MAGPI data, where we exclude spaxels classified as AGN in the definition of the relation but use the relation to measure D4000-based SFRs for all spaxels with no H$\alpha$-based SFRs, including AGN spaxels\footnote{At low S/N, there may be some potential contamination from AGN in the stacking analysis described in this section. However, it is also possible that `retired' (i.e., no longer actively forming stars) regions are falsely classified as AGN \citep[e.g.,][]{stasinska2006, stasinska2008, cidfernandes2010, cidfernandes2011} due to hot post-asymptotic giant branch (AGB) stars and white dwarfs.}. We measure D4000 on a spaxel-by-spaxel basis using the GIST stellar continuum fits, and adopt the broad D4000 definition provided by \citet{hamilton1985} for consistency with low-redshift results (e.g., \citealt{bluck2020a, bluck2020b}).

\begin{figure}
    \centering
    \includegraphics[width=\columnwidth]{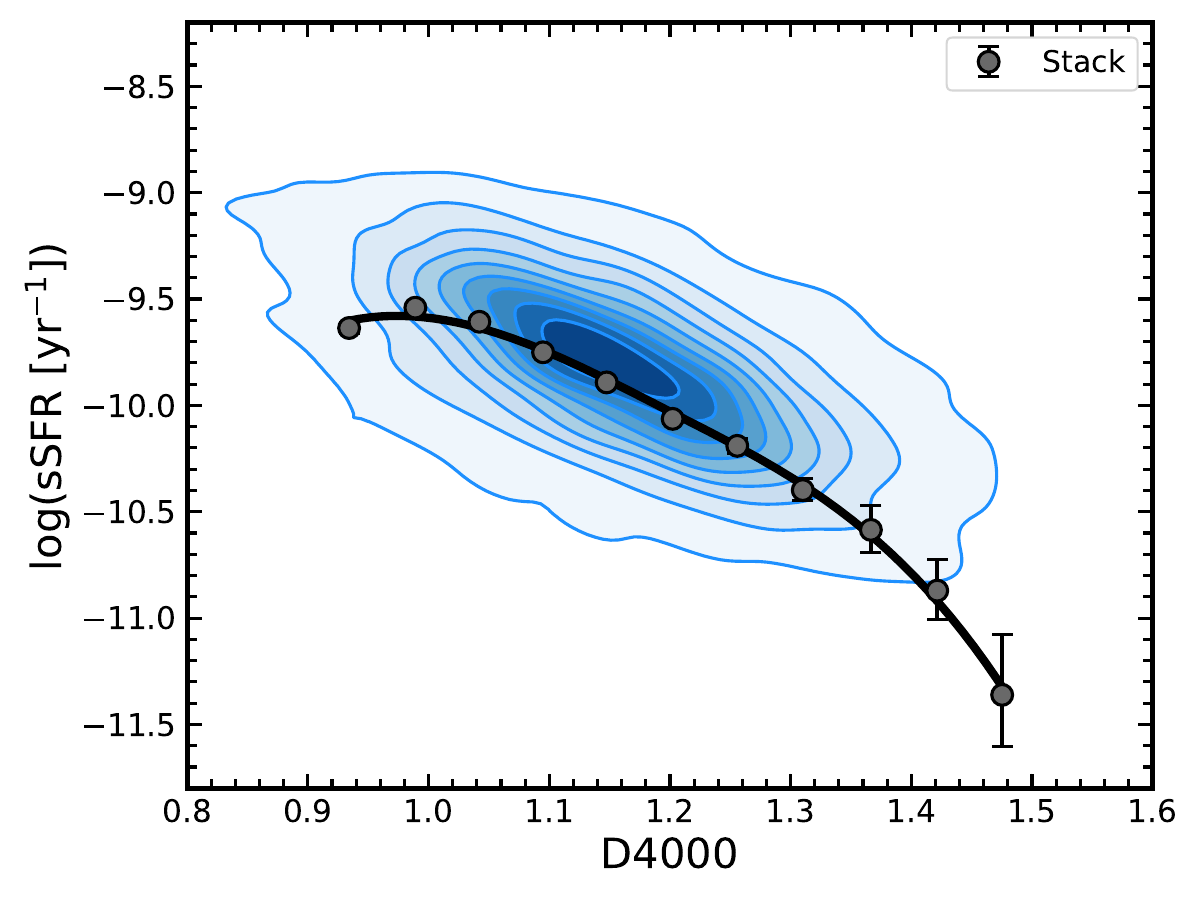}
    \caption{Empirical D4000-sSFR relation measured from stacking spectra in bins of D4000. The measured sSFRs for each D4000 bin are shown by the grey scatter points, overlaid on a one-dimensional smoothing spline fit to the data points shown as a black solid line. The blue contours represent the distribution of sSFRs calculated directly from extinction-corrected \textsc{gist}-derived H$\alpha$ fluxes, plotted for comparison to the stacked points. The stacked points show to be systematically lower than the individual sSFRs, which is due to including weak S/N spectra in the stack, whereas we implemented a S/N $\geq$ 3 criterion for 4 emission lines for the latter (see Section \ref{sec:ha}).}
    \label{fig:d4_ssfr}
\end{figure}

With the \textsc{gist}-derived D4000 values and the reduced spectra, we measure the resolved D4000-sSFR relation by stacking the spectra in a range of D4000 bins to improve the S/N of weak emission lines as shown in Figure~\ref{fig:d4_ssfr}. This is in contrast to previous methods which used individual spaxels for the calibration, such that with stacking we are able to calibrate the relation down to lower sSFRs/higher D4000. We use the following criteria to choose the spaxels/spectra going into the stack: 
\begin{enumerate}
    \item have a measured D4000 with S/N $\geq$ 5,
    \item have a measured M\textsubscript{$\star$} with uncertainties $\leq$ 0.5 dex,
    \item not classified as AGN by the BPT classification scheme.
\end{enumerate}

With regards to the first criterion, we have tested with different S/N cuts on D4000 up to S/N $\geq$ 20 to find that the resulting D4000-sSFR relation does not change significantly. Prior to stacking spectra on a spaxel-by-spaxel basis, we subtract the best-fit continuum measured with \textsc{gist} and de-redshift each spectrum based on their spectroscopic redshifts and velocity field maps. To ensure that all the spectra are matched to an identical range of wavelengths, the de-redshifted spectra are then interpolated (nearest neighbour) to a common grid of rest-frame wavelengths, the range of which is determined from the longest wavelength on the blue side and the shortest wavelength on the red side of all spectra. The spectra are then normalised by stellar mass, such that the unique stellar masses measured for each spaxel are carried through when calculating sSFRs directly from the stacked fluxes. Finally, the spectra are median combined in D4000 bins of fixed width $\mathrm{\delta_{D4000}}$ = 0.05, ranging in values from 0.9 to 1.45. The selected bin width results in each bin having $>$200 spectra with the maximum number of spectra in a bin being $\simeq$2100. We also tested with different bin widths and did not see significant changes in the stacked D4000-sSFR relation. Beyond D4000 $>$ 1.45, the S/N drops dramatically such that we use a wider bin width. We stack spectra for D4000 $>$ 1.45 in 3 bins of 1.45 $<$ D4000 $<$ 1.5, 1.5 $<$ D4000 $\leq$ 1.7, and 1.7 $<$ D4000 $\leq$ 1.9. Despite using a wider bin width resulting in $\gtrsim$6000 spectra per bin for the latter two bins, we are unable to obtain robust fits for H$\alpha$ and H$\beta$ except for the first bin of 1.45 $<$ D4000 $<$ 1.5, resulting in a D4000 limit of $\simeq$1.475. Previous studies such as \citet{bluck2020a} and \citet{thorp2022} have also found that robust SFR measurements for spaxels/galaxies of D4000 $>$ 1.45 were not feasible. Both studies attribute the reason to a steep decline in sSFR at D4000 $>$ 1.4, such that the sSFR varies by $\approx$1 dex over a $\mathrm{\delta_{D4000}}$ of 0.1. However, see \citet{brinchmann2004} and \citet{wang2019} for a continuous derivation at D4000 $> 1.45$. 

After median stacking the spectra in bins of D4000, we fit Gaussian profiles to both the H$\alpha$ and H$\beta$ lines for each stacked spectrum using the code \textsc{lmfit}\footnote{\url{https://lmfit.github.io/lmfit-py/}} \citep{newville2014}. We fit single Gaussian profiles to 4 emission lines: H$\alpha$, [NII]$\lambda\lambda$6549,6584, and H$\beta$. When fitting the emission lines, we take on the following constraints on the fitting parameters: (1) a single line width is fit to all the lines based on the brightest line (i.e., H$\alpha$ for the [NII] doublet), (2) the ratio of [NII]$\lambda$6584/[NII]$\lambda$6549 is fixed to 3.071:1 \citep{sz2000}, and (3) corresponding wavelengths for the emission lines of interest are allowed to vary within $\pm$ 5~\AA \ of the rest-frame wavelengths in vacuum. While the constraint placed on the wavelengths are not strict, the median offsets of the fits are $\simeq$ 0.8~\AA \ and $\simeq$ 0.3~\AA \ for the H$\alpha$ and H$\beta$ lines respectively, suggesting that the constraint of 5~\AA \ is not an issue. We first fit the lines using the Levenberg-Marquardt algorithm, then use the Markov Chain Monte Carlo (MCMC) method to refine our best-fit and uncertainty estimates. The median and 1$\sigma$ values of the posterior probability distributions of the fitted parameters are then used to measure integrated fluxes and their errors. The integrated H$\alpha$ fluxes for each bin are then dust-corrected for both foreground and intrinsic extinction via the same method outlined in Section \ref{sec:ha}, using the integrated H$\beta$ fluxes to measure the Balmer decrement. Foreground extinction is corrected for by using the median $E(B - V)$ of the stack. We then use the median spectroscopic redshift of each bin to convert the extinction-corrected H$\alpha$ fluxes to luminosities, and then to sSFRs. 

The resulting resolved D4000-sSFR relation for 0.25 $\leq z \leq$ 0.424 is shown in Figure \ref{fig:d4_ssfr}. The blue contours represent the distribution of spaxels with measurements of both H$\alpha$-derived SFRs and D4000 (7,788 spaxels), whereas the grey circles along with the black solid line indicate the stacked D4000-sSFR relation. The black line is given by a one-dimensional smoothing spline fit to the data points. To convert D4000 to SFRs, each spaxel with a measured D4000 is first matched to the closest value in D4000 on the D4000-sSFR relation. Then, a sample of possible sSFR values are generated based on the input D4000 and its error via Monte Carlo, where the median of the distribution is taken as the output sSFR. The median absolute deviation (MAD) standard deviation of the sSFR distribution and the standard deviation of the scatter in the recovery of SFRs with D4000 (see Figure \ref{fig:d4hacomp}) are added in quadrature to obtain uncertainties on the sSFRs. The latter uncertainty is measured from comparing $\mathrm{\Sigma_{SFR, D4000}}$ and $\mathrm{\Sigma_{SFR, H\alpha}}$ for the same spaxels, where the standard deviation of $\mathrm{\Delta \log}$($\mathrm{\Sigma_{SFR}}$) [$\equiv$ $\mathrm{\log}$($\mathrm{\Sigma_{SFR, D4000}}$) - $\mathrm{\log}$($\mathrm{\Sigma_{SFR, H\alpha}}$)] is added in quadrature to reflect how well $\mathrm{\Sigma_{SFR, H\alpha}}$ measurements are recovered with D4000. Additional discussion along with tests performed on the reliability of D4000 as a SFR indicator are presented in Appendix \ref{sec:app1}. The chosen sSFR is then converted to a $\mathrm{\Sigma_{SFR}}$ using the following equation: 
\begin{equation} \label{eq:4}
    \mathrm{\Sigma_{SFR, D4000} \ (M_{\odot} \ yr^{-1} \ kpc^{-2})} = \mathrm{sSFR} \times \mathrm{\Sigma_{\star}},
\end{equation}
where the units of $\mathrm{\Sigma_{\star}}$ are in M\textsubscript{\sun} kpc\textsuperscript{-2}. SFRs are measured in the same manner, except in units of M\textsubscript{\sun} yr\textsuperscript{-1}. Total SFRs are then measured by summing up all the spaxels with measured D4000-derived SFRs that lie in the dilated masks, similarly done with the H$\alpha$ SFRs. 

\subsection{Global and local star formation metrics}
\label{sec:metrics}
To compare different subsamples of galaxies in our analysis, we consider the known dependencies of SFRs on other properties (e.g., \citealt{gonzalezdelgado2016}), most notably stellar mass. Similar in definition to the commonly measured sSFR, we choose to use $\mathrm{\Delta \Sigma_{SFR}}$ and $\Delta$SFR parameters (e.g., \citealt{ellison2018, bluck2020a, bluck2020b}). The parameters are defined as the following:
\begin{align} 
    \Delta \mathrm{\Sigma_{SFR}} &= \log_{10}(\mathrm{\Sigma_{SFR, spax}}) - \log_{10}(\mathrm{\Sigma_{SFR, MS}}) \label{eq:5} \\
    \Delta \mathrm{SFR} &= \log_{10}(\mathrm{SFR_{gal}}) - \log_{10}(\mathrm{SFR_{MS}}) \label{eq:6} 
\end{align}
where both parameters describe the logarithmic offsets from the resolved and global SFMS, respectively. The parameters are defined such that any galaxy/spaxel above the main sequence has positive $\Delta$-parameter indicating higher levels of star formation activity than a galaxy/spaxel on the main sequence, and vice versa. While $\Delta$SFR measures how star-forming a galaxy is overall for its stellar mass, $\mathrm{\Delta \Sigma_{SFR}}$ is defined in a similar way except that it measures the level of star formation activity on a local scale equivalent to that of a MUSE spaxel. 

\begin{figure}
    \centering
    \includegraphics[width=\columnwidth]{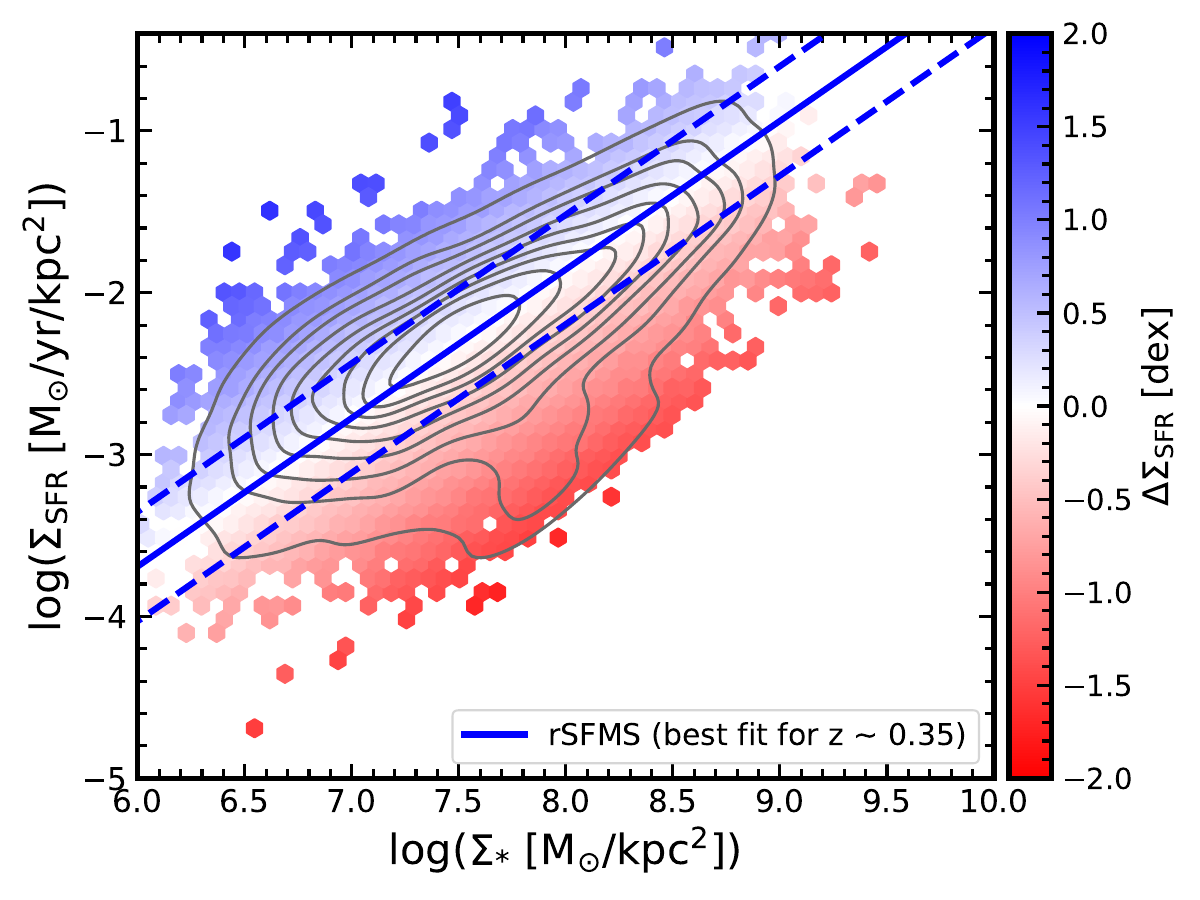}
    \caption{The resolved SFMS relation measured between $\mathrm{\Sigma_{SFR}}$ and $\mathrm{\Sigma_{\star}}$ for 0.25 $\leq z \leq$ 0.424. The blue solid line represents the resolved SFMS measured with \textsc{ltsfit}, with the blue dashed lines representing the root-mean-square error of the fit. The hexagonal bins are colour-coded by $\Delta \mathrm{\Sigma_{SFR}}$, the logarithmic offset from the resolved SFMS.}
    \label{fig:rsfms}
\end{figure}

Given the definitions of the $\Delta$-parameters, we measure both the global and resolved star-forming main sequences. We assume that the dependence of SFR on stellar mass can be well described by a single power-law such that the SFMS is a linear relation in log-log space. We use the code \textsc{ltsfit}\footnote{\url{https://pypi.org/project/ltsfit/}} \citep{cappellari2013} to measure a best linear fit to the SFMS. 

To define the resolved SFMS, we only use the spaxels with H$\alpha$-based $\mathrm{\Sigma_{SFR}}$ (total of 12,198 spaxels) measurements to ensure that the resolved SFMS best represents a population of star-forming spaxels. We show the resulting fit along with all the spaxels with measured $\mathrm{\Sigma_{SFR}}$ and $\mathrm{\Sigma_{\star}}$ (including D4000-derived $\mathrm{\Sigma_{SFR}}$ brings the total up to 13,504 spaxels) in Figure \ref{fig:rsfms}, where each hexagonal bin is colour-coded by $\Delta \mathrm{\Sigma_{SFR}}$ as defined by Equation \ref{eq:5}. The best-fit equation is given below:
\begin{equation} \label{eq:7}
    \log_{10}(\mathrm{\Sigma_{SFR}}) [\mathrm{M_{\odot} \ yr^{-1} \ kpc^{-2}}] = 0.918 \times \log_{10}(\mathrm{\Sigma_{\star}}) - 9.196,
\end{equation}
with the root-mean-square error on the fit being $\sim$0.338 dex, shown by the blue dashed line in Figure \ref{fig:rsfms}. The errors on the slope and intercept are 0.005 and 0.006, respectively. The global SFMS is defined similarly such that we only use galaxies with SFR\textsubscript{H$\alpha$} (total of 211 galaxies) for the fit. We show the resulting global SFMS fit, along with the distribution of a total of 302 galaxies with measured SFR (including D4000-based) and M\textsubscript{$\star$} in Figure \ref{fig:gsfms}. The equation for the global SFMS is given below:
\begin{equation} \label{eq:8}
    \log_{10}(\mathrm{SFR}) [\mathrm{M_{\odot} \ yr^{-1}}] = 0.748 \times \log_{10}(\mathrm{M_{\star}}) - 7.726,
\end{equation}
with the root-mean-square error on the fit being $\sim$0.551 dex, and the slope and intercept errors being 0.038 and 0.055, respectively. 

\begin{figure}
    \centering
    \includegraphics[width=\columnwidth]{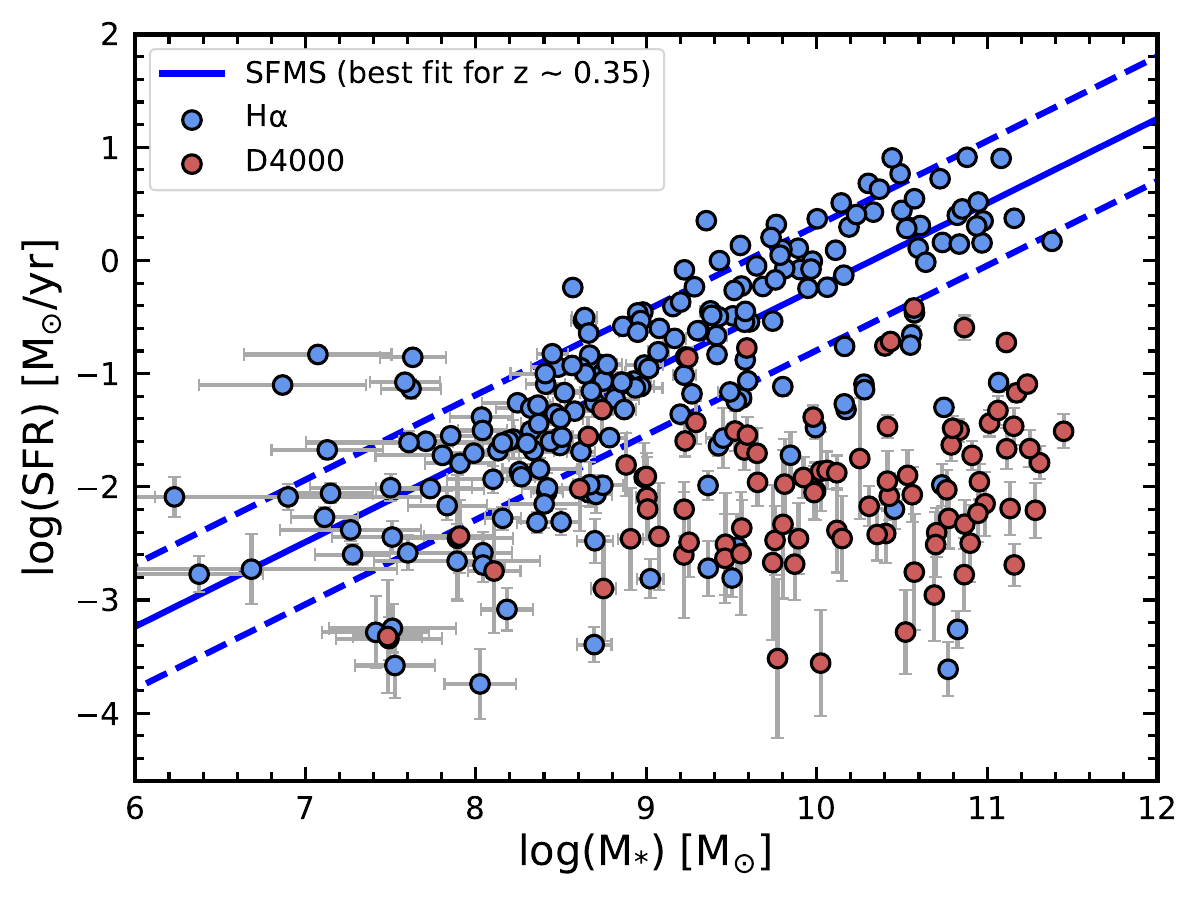}
    \caption{The global SFMS relation measured between SFR and M\textsubscript{$\star$} for 0.25 $\leq z \leq$ 0.424. The blue lines are defined analogously to those shown in Figure \ref{fig:rsfms}. Error bars on the measurements are shown in grey. Blue and red points indicate galaxies with H$\alpha$- and D4000-based SFRs, respectively.}
    \label{fig:gsfms}
\end{figure}

\begin{figure}
    \centering
    \includegraphics[width=\columnwidth]{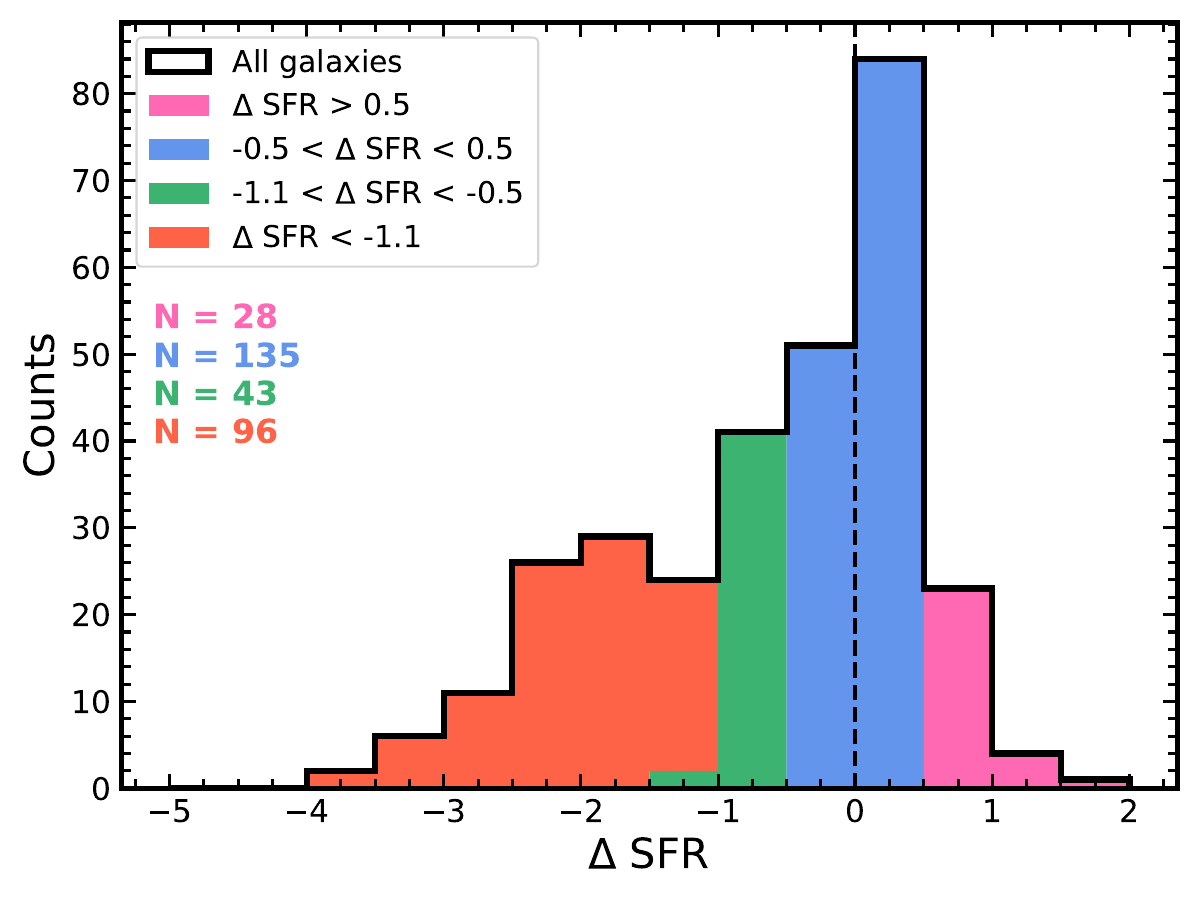}
    \caption{Histogram showing the distribution of $\Delta$SFR values of a total of 302 MAGPI galaxies, colour-coded by different global SF states according to $\Delta$SFR. Sample size per global SF state is listed below the legend, following the same colour code.}
    \label{fig:dsfrhist}
\end{figure}

We define global SF states as a function of logarithmic offset from the SFMS using the following definitions to classify galaxies in our sample:  
\begin{enumerate}
    \item Above the SFMS: $\Delta$SFR $>$ 0.5 dex,
    \item SFMS: -0.5 dex $<$ $\Delta$SFR $<$ 0.5 dex,
    \item Just below the SFMS: -1.1 dex $<$ $\Delta$SFR $<$ -0.5 dex,
    \item Far below the SFMS: $\Delta$SFR $<$ -1.1 dex.
\end{enumerate}
These classifications are comparable to those used for the MANGA sample, allowing a fair comparison with results based on the local Universe \citep{bluck2020a, bluck2020b}. The distribution of $\Delta$SFR values for our sample are shown in Figure \ref{fig:dsfrhist}, colour-coded by the different SF states as defined above. These classifications result in a total of 301 galaxies. Finally, 8 galaxies are removed from the sample that have a global measurement of SFR and stellar mass but individually do not have spaxels that meet our S/N criteria in both the stellar mass and SFR maps simultaneously. This results in our final sample of 294, such that we have 27 galaxies in the $\Delta$SFR $>$ 0.5 bin, 131 galaxies in the main sequence bin, 42 galaxies in the -1.1 $<$ $\Delta$SFR $<$ -0.5 bin, and 94 galaxies in the least star-forming bin.  

\subsection{Derivation of radial profiles}
\label{sec:radprof}
Following the definitions of $\mathrm{\Delta \Sigma_{SFR}}$ and $\Delta$SFR, we construct radial profiles of star formation activity as a function of global SF state to probe the physical mechanisms at play within MAGPI galaxies. We apply the same methodology towards measuring radial profiles in Age\textsubscript{L} as well, where stellar ages are derived based on stellar continuum fits using the C3K spectral templates (Section \ref{sec:elinefit}). We use the structural parameters derived with \textsc{ProFound}, which include effective radii (i.e., $R_{\rm e}$), position angles, and axial ratios, to define radial bins of elliptical annuli. These parameters were measured using $i$-band MUSE images reconstructed with the SDSS $i$-band filter transmission curve. The structural parameters given by \textsc{ProFound} are PSF-convolved measurements, as \textsc{ProFound} does not take either the size or shape of the PSF into account in source detection. 

Fixing the locations of peak flux in the white light image as the galaxy centre, we calculate galactocentric distances for each spaxel in units of $R_{\rm e}$ for each galaxy. We choose to use fixed discrete bins of width 0.5 $R_{\rm e}$ from 0 to 3.0 $R_{\rm e}$ to bin values of $\mathrm{\Delta \Sigma_{SFR}}$ from all galaxies with measured $\mathrm{\Delta \Sigma_{SFR}}$ maps. The radial extent out to which we probe is set by the number of galaxies and spaxels contributing to each radial bin, such that only bins with at least 10 galaxies ($\mathrm{N_{gal}} \geq$ 10) and 50 spaxels ($\mathrm{N_{spax}} \geq$ 50) were included. The cuts are chosen to exclude bins with low number statistics in either the number of galaxies or spaxels. Details on the total number of galaxies and spaxels contributing to each bin for both $\mathrm{\Delta \Sigma_{SFR}}$ and Age\textsubscript{L} profiles are given in Appendix \ref{sec:app2}. We confirm that our results are not significantly affected by the choice of fixed versus smoothed binning approaches or bin width, consistent with \citet{bluck2020b}. To measure a population-averaged radial profile, we take the median of all spaxels belonging to all galaxies within each global SF state for each radial bin. We then measure the standard error on the median $\mathrm{\Delta \Sigma_{SFR}}$ with bootstrap. 

\begin{figure}
    \centering
    \includegraphics[width=\columnwidth]{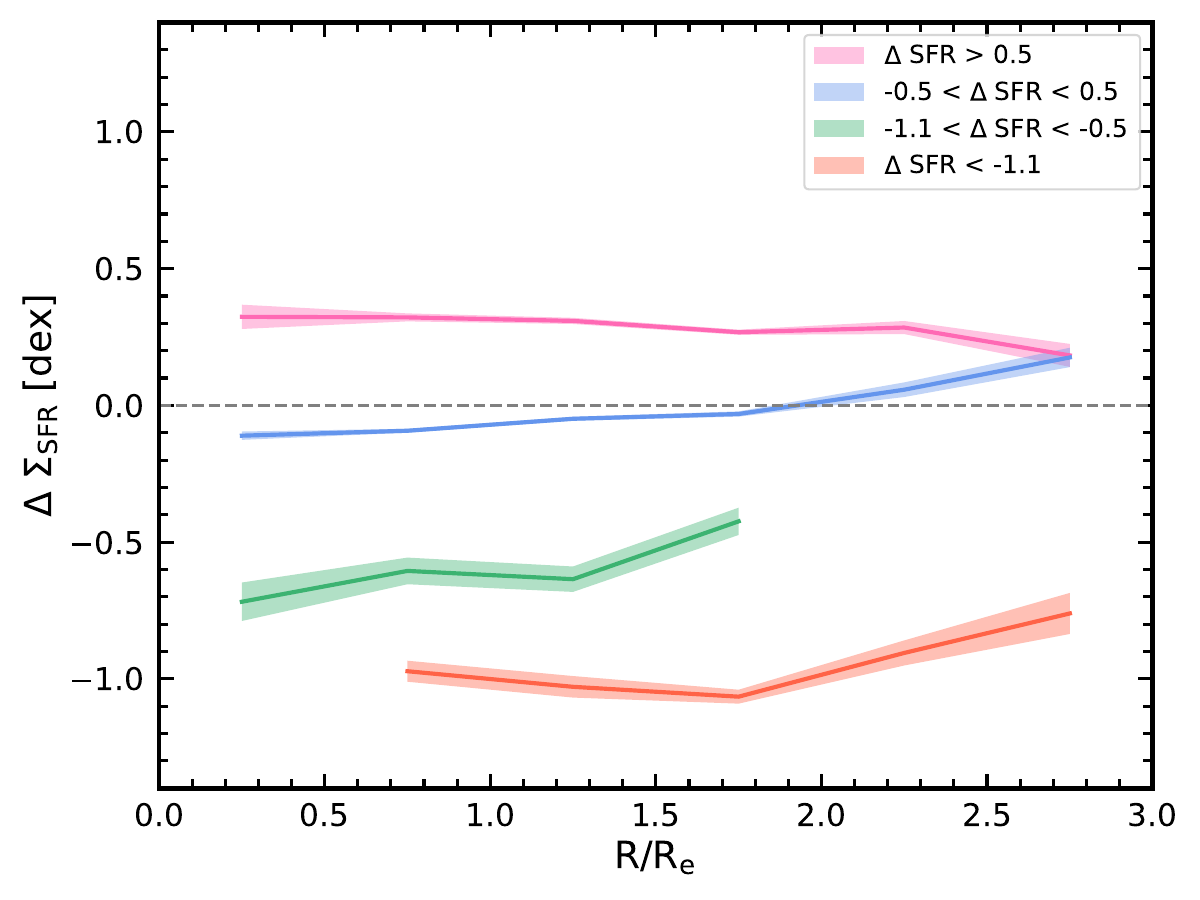}
    \caption{Population-averaged (i.e., median) radial $\mathrm{\Delta \Sigma_{SFR}}$ profiles for 4 global SF states as defined by different bins of $\Delta$SFR, i.e., different ranges of locations with respect to the global SFMS. A grey dashed line at $\mathrm{\Delta \Sigma_{SFR}}$ = 0 is shown to indicate the value at which a spaxel is as star-forming as a spaxel of the same $\mathrm{\Sigma_{*}}$ located on the resolved SFMS. Shaded regions for each profile denote the bootstrap errors based on the median.}
    \label{fig:dsfrrp}
\end{figure}

Due to our relatively small sample we do not apply an inclination cut. Using a larger sample, \citet{bluck2020b} have shown that their radial profiles are robust against restrictions to face-on galaxies or adopting Euclidean distance formula rather than elliptical aperture bins consistent with \citet{ellison2018}. 

\begin{figure}
    \centering
    \includegraphics[width=\columnwidth]{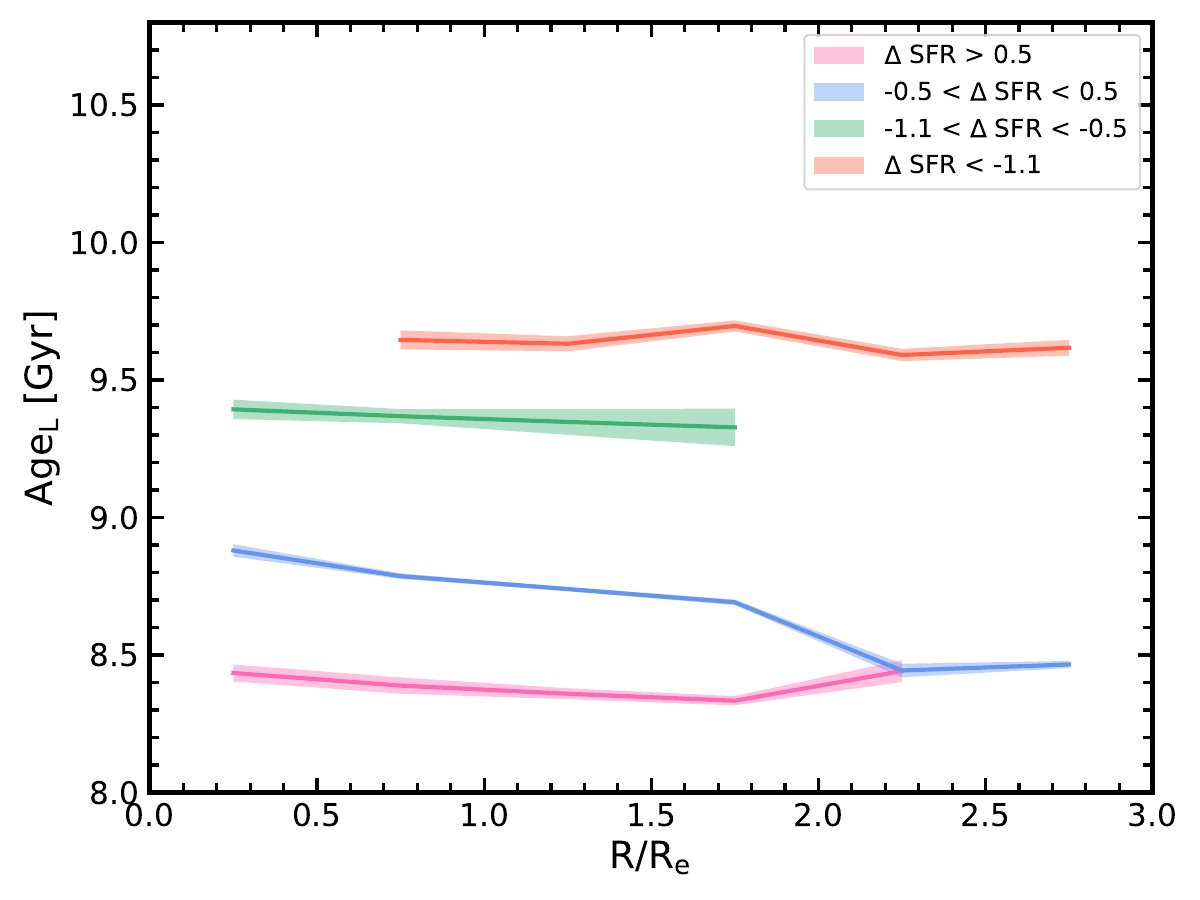}
    \caption{Radial profiles of the luminosity-weighted stellar ages for each of the defined global SF states, with the colour-coding identical to that of Figure \ref{fig:dsfrrp}. As done with the $\mathrm{\Delta \Sigma_{SFR}}$ profiles, the uncertainties on the median stellar age per radial bin are measured with bootstrap.}
    \label{fig:agerp}
\end{figure}

\section{Results}
\label{sec:results}
In this section, we explore radial trends in both star formation rates and stellar ages as a function of global SF state, to understand what potential processes govern our sample of MAGPI galaxies at $z \sim$ 0.3. We compare the radial trends in a qualitative and quantitative manner, where for the latter we measure radial slopes with Monte Carlo derived uncertainties by making linear fits to the profiles. The slopes and errors are shown in Table \ref{tab:rpslopes}. We show the $\mathrm{\Delta \Sigma_{SFR}}$ profiles in Section \ref{sec:dsfrprof} and put them in context with the Age\textsubscript{L} profiles in Section \ref{sec:agelum}. 

\subsection{\texorpdfstring{$\boldsymbol{\mathrm{\Delta \Sigma_{SFR}}}$}{deltasigmasfr} profiles} \label{sec:dsfrprof}
The $\mathrm{\Delta \Sigma_{SFR}}$ radial profiles are shown in Figure \ref{fig:dsfrrp}, colour-coded in the same manner as shown in Figure \ref{fig:dsfrhist}. For the $\Delta$SFR $>$ 0.5 bin (pink), galaxies exhibit enhanced star formation activity with respect to the resolved SFMS (i.e., positive $\mathrm{\Delta \Sigma_{SFR}}$), where the overall trend is flat (slope = -0.05 $\pm$ 0.01 dex $R_{\rm e}$\textsuperscript{-1}) across all radial distances probed. The flatness of the profile suggests that the physical processes at play do not enhance specific regions in galaxies above the SFMS at $z \sim 0.3$. For all galaxies in this bin, all spaxels have H$\alpha$-measured $\mathrm{\Sigma_{SFR}}$ - indicating that the star formation probed in these galaxies is recent. 

For galaxies on the main sequence (i.e., $-0.5 < \Delta$ SFR $< 0.5$; blue), the profile shows a slightly positive gradient with a measured slope of 0.11 $\pm <$0.01 dex $R_{\rm e}$\textsuperscript{-1}. We find 3 galaxies with notably high spaxel contribution fractions to be responsible for the small enhancement at the outskirts ($R \gtrsim$ 2 $R_{\rm e}$). Two of these galaxies are among the more highly star-forming galaxies in this bin ($\Delta$SFR $\simeq$ 0.42 and 0.49 dex), and all 3 of them exhibit on average higher $\mathrm{\Sigma_{SFR}}$ with increasing galactocentric radii. While it is not trivial to interpret this as a sign of inside-out quenching and/or preferential enhancement of SF in the outskirts, we are able to confirm that the uptick in the outskirts is statistically robust. Removing these three galaxies results in a flatter profile but drops the last radial bin out to 3 $R_{\rm e}$ below our criteria of a minimum of 10 galaxies. It is also noteworthy that this uptick has not been observed in previous studies, such as those based on the local Universe \citep{ellison2018, bluck2020b}, showcasing the strength of MAGPI in probing the far outskirts of galaxies. The inner flat (slope = 0.06 $\pm$ 0.01 dex $R_{\rm e}$\textsuperscript{-1} for $R \leq 1.5 R_{e}$; refer to Table \ref{tab:mangaslopes}) profile suggests that if there are any competing physical mechanisms at play, they balance each other such that the level of star formation activity stays fairly uniform from the galaxy centre towards the outskirts. Despite the galaxies in this bin lying on or near the SFMS, the majority of the average radial profile overall sits below $\mathrm{\Delta \Sigma_{SFR}}$ = 0 (grey dashed line). This suggests that what is observed in the $\mathrm{\Delta \Sigma_{SFR}}$ does not necessarily align with what is suggested by a galaxy's $\Delta$SFR, such that it is important to investigate galaxy-wide star formation activity on both local and global scales. Similar results have been observed with local (e.g., MaNGA; \citealt{ellison2018}) galaxies, in particular, satellite galaxies on the main sequence exhibit a flat $\mathrm{\Delta \Sigma_{SFR}}$ profile at $\mathrm{\Delta \Sigma_{SFR}} \sim -0.1$ at all radial distances probed up to 1.5 $R_{\rm e}$ \citep{bluck2020b}. 

We observe a positive (slope = 0.17 $\pm$ 0.02 dex $R_{\rm e}$\textsuperscript{-1}) radial gradient in $\mathrm{\Delta \Sigma_{SFR}}$ out to 2 $R_{\rm e}$ within the -1.1 $< \Delta$SFR $<$ -0.5 bin (green). The entire profile is located at $\mathrm{\Delta \Sigma_{SFR}} \sim -0.65$, which together with the positive gradient, is suggestive of inside-out quenching. Moreover, the contribution of H$\alpha$-detected spaxels begins to decrease to $\sim40$ per cent for galaxies in this bin and to $<10$ per cent in the $\Delta$SFR $<$ -1.1 bin (red), further supporting the onset of quenching (Appendix~\ref{sec:app1}). In contrast to what is measured for the $\Delta$SFR $>$ 0.5 (pink) and -0.5 $< \Delta$SFR $<$ 0.5 (blue) bins, we are only able to robustly measure the profile out to 2 $R_{\rm e}$. For $\sim$34 per cent of spaxels in this bin, we only measure an upper limit on $\mathrm{\Delta \Sigma_{SFR}}$ because they have D4000 $>$ 1.475. As we do not include these upper limits into consideration for the profile measurement, the profile shown in Figure~\ref{fig:dsfrrp} is an upper limit (Appendix~\ref{sec:app1}).

In our lowest $\Delta$SFR bin (red), we observe an overall positive gradient with a slope of 0.11 $\pm$ 0.01 dex $R_{\rm e}$\textsuperscript{-1}, again suggestive of inside-out quenching. These results are similar to that of the -1.1 $< \Delta$SFR $<$ -0.5 bin (green), with the exception that galaxies in this bin exhibit lower values of $\mathrm{\Delta \Sigma_{SFR}}$ across all radial bins probed. Moreover, the contribution of H$\alpha$-detected spaxels is far lower in this bin, reaching an average of $\sim$10 per cent per radial bin. As such, galaxies in this bin are also subject to the same limitations imposed by the measured D4000-sSFR relation, where the fraction of spaxels with upper limits are high towards the galaxy centre ($\sim$97 per cent in the 0 -- 0.5 $R_{\rm e}$ bin, compared to $\sim$84 percent in the 2 -- 2.5 $R_{\rm e}$ bin). This is expected given that older stellar populations (and as such larger D4000) likely reside in galaxy centres. Given that quite a few galaxies have measured D4000 values higher than that limit, we are potentially biasing our sample of `quenched' galaxies to those with some residual star formation. We explore this further in Section \ref{sec:disc} and Appendix \ref{sec:app1}. 

\subsection{Luminosity-weighted stellar ages}
\label{sec:agelum}
We show the luminosity-weighted stellar age (Age\textsubscript{L}) profiles, following the same methodology of stellar continuum fitting with \textsc{gist} as described in Section \ref{sec:elinefit}, in Figure \ref{fig:agerp}. Starting with the highest star-forming bin (i.e., $\Delta$SFR $>$ 0.5; pink), we observe an overall uniform distribution of stellar ages with a slope of -0.01 $\pm$ 0.01 Gyr $R_{\rm e}$\textsuperscript{-1}. Galaxies in this bin consist of the youngest stellar ages out of all global SF states, which is in line with what we observe with the $\mathrm{\Delta \Sigma_{SFR}}$ profile. 

\begin{table}
    \centering
    \caption{Table of measured slopes of $\mathrm{\Delta \Sigma_{SFR}}$ and Age\textsubscript{L} profiles shown in Figures \ref{fig:dsfrrp} and \ref{fig:agerp}.}
    \begin{tabular}{ccc}
    \hline
     Global SF state & $\mathrm{\Delta \Sigma_{SFR}}$ & Age\textsubscript{L} \\
     (dex) & (dex $R_{\rm e}$\textsuperscript{-1}) & (Gyr $R_{\rm e}$\textsuperscript{-1}) \\ \hline
     $\Delta$SFR $>$ 0.5 & -0.05 $\pm$ 0.01 & -0.01 $\pm$ 0.01 \\
     -0.5 $<$ $\Delta$SFR $<$ 0.5 & 0.11 $\pm <$0.01 & -0.18 $\pm <$0.01 \\
     -1.1 $<$ $\Delta$SFR $<$ -0.5 & 0.17 $\pm$ 0.02 & -0.04 $\pm$ 0.02 \\
     $\Delta$SFR $<$ -1.1 & 0.11 $\pm$ 0.01 & -0.02 $\pm$ 0.01 \\
    \hline
    \end{tabular}
    \label{tab:rpslopes}
\end{table}

On the other hand, we observe a negative gradient (slope = -0.18 $\pm <$ 0.01 Gyr $R_{\rm e}$\textsuperscript{-1}) in the stellar ages for the main sequence (i.e., -0.5 $<$ $\Delta$SFR $<$ 0.5; blue) population, consistent with inside-out growth \citep{munoz-mateos07, vandokkum2010}. This is in contrast to the slightly positive gradient we observe out to 2.5 $R\mathrm{e}$ in $\mathrm{\Delta \Sigma_{SFR}}$ (slope = 0.11 $\pm <$ 0.01 dex $R_{\rm e}$\textsuperscript{-1}), where we may be capturing lower levels of recent SF and older stellar populations in the centre. 

For the -1.1 $< \Delta$SFR $<$ -0.5 bin (green), we observe a flat profile with a slope of -0.04 $\pm$ 0.02 Gyr $R_{\rm e}$\textsuperscript{-1}. This is in contrast to the positive gradient observed in $\mathrm{\Delta \Sigma_{SFR}}$, where the galaxy centre is less star-forming than the outskirts. Given that the $\mathrm{\Delta \Sigma_{SFR}}$ profile arises from a combination of H$\alpha$ and D4000, any residual SF probed by H$\alpha$ is not being captured in the Age\textsubscript{L} profile likely due to timescale differences ($\lesssim$10~Myr for H$\alpha$ vs. $\lesssim$1~Gyr for the stellar continuum). 

A flat profile is observed in the $\Delta$SFR $<$ -1.1 (red) bin as well (slope = -0.02 $\pm$ 0.01 Gyr $R_{\rm e}$\textsuperscript{-1}), where the oldest ages out of all 4 global SF states are observed. We observe similar trends with the -1.1 $< \Delta$SFR $<$ -0.5 bin, where in addition to a flat Age\textsubscript{L} profile, we observe a positive gradient in $\mathrm{\Delta \Sigma_{SFR}}$ despite the significantly decreased fraction of H$\alpha$ spaxels in this bin. The difference in radial trends may also stem from the use of Voronoi binning for stellar continuum fitting. As Voronoi binning is not applied to the emission line fluxes or D4000 measurements, the spatial variation we are able to probe is not as fine with the ages. Nevertheless, the overall increase in range of stellar ages with decreasing $\Delta$SFR is in line with the expectation of galaxies hosting older stellar populations with decreased SFRs. 

\section{Discussion}
\label{sec:disc}
We have investigated the $\mathrm{\Delta \Sigma_{SFR}}$ and luminosity-weighted stellar ages of galaxies across different regions of the SFMS within the MAGPI sample. In this section, we first address the probability of a potential redshift evolution of galaxy properties among local ($z \sim$ 0), intermediate ($z \sim$ 0.3; this study), and high ($z \sim$ 2) redshift regimes. We then focus on understanding the potential physical mechanisms at play while also connecting our results to those of other similar IFS-based studies. We also put our results in context with what has been found with cosmological simulations. 

\subsection{Do we observe an evolution in star formation activity with redshift at \texorpdfstring{$z \boldsymbol{\sim}$}{sim} 0.3?}
\label{sec:zevo}
Galaxy evolution studies indicate that mass-averaged galaxy populations have evolved significantly from $z\sim 2$ to the present day (e.g., \citealt{fw2020}; and references therein). These changes are likely driven by the evolution of accretion, environment, and the build up of passive galaxies through quenching. Here we explore our $\mathrm{\Delta \Sigma_{SFR}}$ profiles in the context of this evolution.

\begin{figure}
    \centering
    \includegraphics[width=\columnwidth]{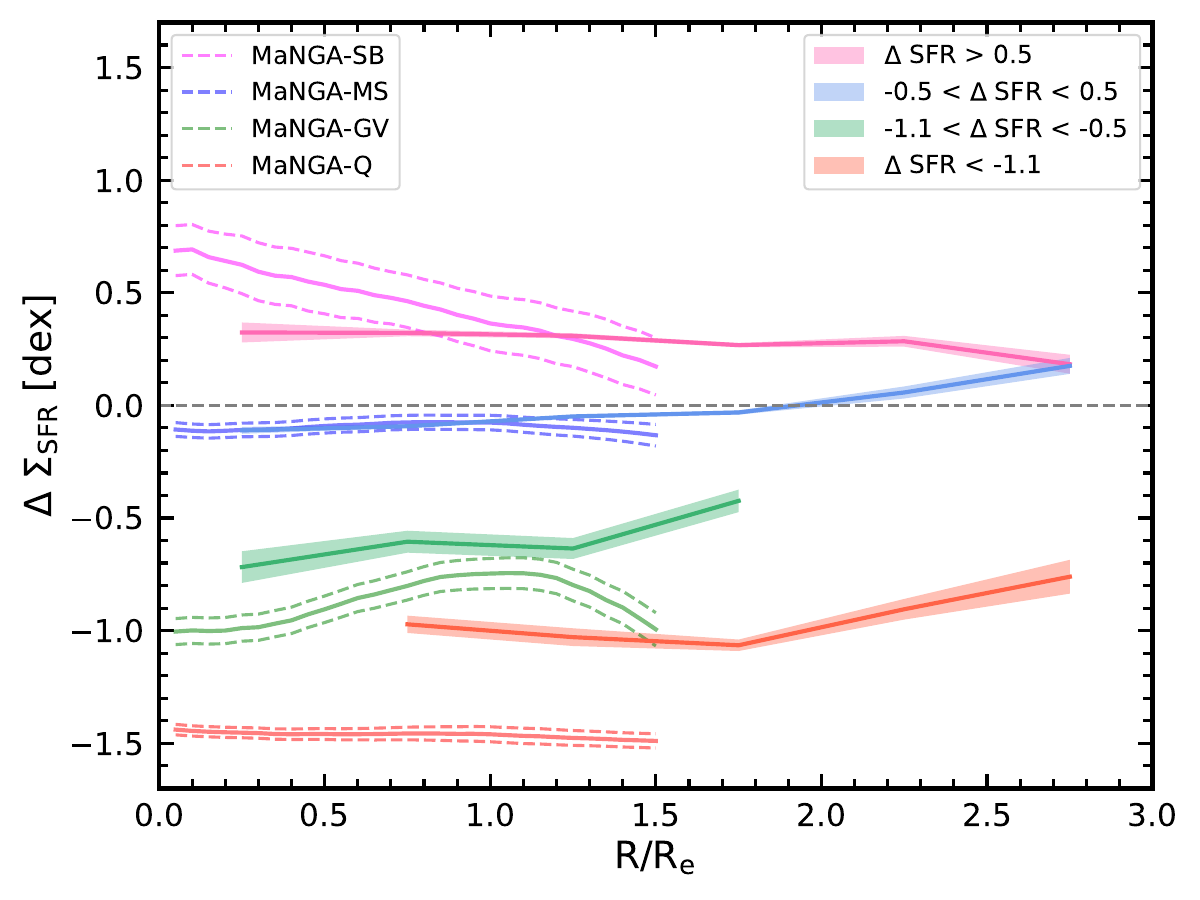}
    \caption{Overlay of $\mathrm{\Delta \Sigma_{SFR}}$ radial profiles of satellite galaxies from \citet{bluck2020b} on profiles shown in Figure \ref{fig:dsfrrp}. The colour-coding is identical between the two profiles. The binning method used in \citet{bluck2020b} is of the smoothed binning approach, where bins are chosen to overlap in range with neighbouring bins.}
    \label{fig:bluck20bcomp}
\end{figure}

Starting with $z \sim$ 0, we overplot the radial profiles from MaNGA data (solid and dashed lines; \citealt{bluck2020b}) on top of our results in Figure \ref{fig:bluck20bcomp} to examine any differences in $\mathrm{\Delta \Sigma_{SFR}}$. We measure slopes for the MaNGA and MAGPI profiles, the latter for which we re-measure the slopes only up to 1.5 $R_{\rm e}$ for direct comparison to MaNGA. The slopes are given in Table \ref{tab:mangaslopes}. Given that the majority of our MAGPI sample ($\sim$76 per cent) belong to groups identified using the algorithm introduced by \citet{robotham2011}, we plot MaNGA radial profiles for satellites only for fairer comparison. Aside from the main sequence (i.e., -0.5 $< \Delta$SFR $<$ 0.5; blue) profile, all star-forming bins show distinctive profiles in overall shape and/or normalisation of $\mathrm{\Delta \Sigma_{SFR}}$ values. 

Notably, the $\Delta$SFR $>$ 0.5 bin (pink) as measured with MaNGA shows a strong negative gradient (slope = -0.34 $\pm$ 0.02 dex $R_{\rm e}$\textsuperscript{-1}) in star formation, whereas the profile measured with MAGPI is flat across the same radial distances probed (slope = -0.02 $\pm$ 0.02 dex $R_{\rm e}$\textsuperscript{-1} for $R \leq 1.5 R_{\rm e}$). The negative gradient has been observed in star-forming galaxies in other MaNGA-based studies \citep{ellison2018, wang2019}, which seems to suggest that this is a common trend in the local Universe. In particular, \citet{wang2019} find this central enhancement to become more distinct with increasing stellar mass. A similar trend is seen with SAMI galaxies \citep{medling2018}, showing centrally concentrated SF in galaxies located $>1\sigma$ above the main sequence (however, we note that this study used $\mathrm{\Sigma_{SFR}}$ and not sSFR, where the latter is a much more comparable parameter to $\mathrm{\Delta \Sigma_{SFR}}$). In contrast, \cite{bluck2020b} find a flattening with increasing stellar mass for the above MS profile with $\log{\mathrm{M_{\star}}}=10.5-11$ galaxies being consistent with a flat profile. While we do not split our sample in stellar mass bins due to low number statistics, we note that the majority of MAGPI galaxies above the MS are in the mass ranges of $\log{\mathrm{M_{\star}}}=9-9.5$ or lower (Figure~\ref{fig:gsfms}), therefore implying evolution within this sample.

\begin{figure*}
    \centering
    \includegraphics[scale=0.45]{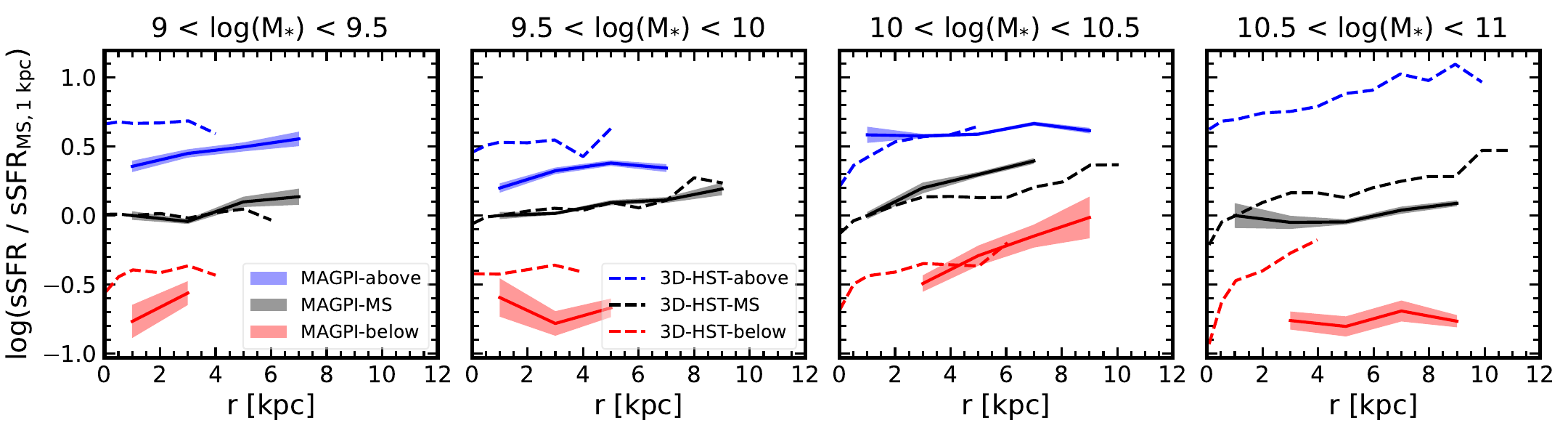}
    \caption{sSFR profiles of MAGPI galaxies (solid lines with shaded regions) categorized into 3 groups with respect to the SFMS, as adopted with 3D-\textit{HST} data in \citet{nelson2016}, where blue denotes galaxies above the MS, black denotes galaxies on the MS, and red the galaxies below the MS. The sSFR profiles from \citet{nelson2016}, representative of galaxies at $z \sim$ 1, are also shown as dashed lines with the same colours. All profiles have been normalised by the sSFR value of the respective MS (black) profile at $r \sim$ 1 kpc.}
    \label{fig:nelson16comp}
\end{figure*}

At $z\sim1$, \citet{nelson2016} find overall flat H$\alpha$ surface brightness profiles when normalised by the SFMS, such that SF is enhanced everywhere in galaxies above the SFMS. However, they also find increasingly centrally \textit{depressed} profiles in sSFR and H$\alpha$ equivalent width with increasing stellar mass. For a more direct comparison, we re-measure the radial profiles in sSFR ($\equiv \mathrm{\Sigma_{SFR} / \Sigma_{\star}}$) in 3 different star-forming bins, across 4 bins of stellar mass, as done in \citet{nelson2016}. We show the definitions of the star-forming bins below, where the upper bound on the `above the SFMS' bin and the lower bound on the `below the SFMS' bin has been modified to include the rest of the MAGPI sample.
\begin{enumerate}
    \item Above the SFMS: $\Delta$SFR $>$ 0.4 dex,
    \item SFMS: -0.4 dex $<$ $\Delta$SFR $\leq$ 0.4 dex,
    \item Below the SFMS: $\Delta$SFR $\leq$ -0.4 dex,
\end{enumerate}
Similarly done with the $\mathrm{\Delta \Sigma_{SFR}}$ profiles, we impose galaxy and spaxel cuts such that only radial bins with at least 5 galaxies (N\textsubscript{gal} $\geq$ 5) and 25 spaxels (N\textsubscript{spax} $\geq$ 25) are included. The cuts imposed are less stringent here (see Section \ref{sec:radprof}) due to reduced sample statistics coming from binning the profiles into multiple stellar mass regimes. The profiles, along with those from \citet{nelson2016} (bottom panel of Figure 14 in their study), are shown in Figure \ref{fig:nelson16comp}. We have normalised all profiles by the measured sSFR value of the SFMS profile at $r \approx$ 1 kpc, to facilitate easier comparison. Starting with galaxies in the lowest mass bin, we observe overall flat profiles for galaxies above (blue; slope = 0.03 $\pm <$0.01 dex kpc\textsuperscript{-1}) the SFMS, in line with what is observed with the 3D-\textit{HST} results \citep{nelson2016}. However, with increasing stellar mass, we do not observe the same trend of increasingly centrally depressed profiles; we measure similar slopes of 0.02 $\pm <$0.01 and 0.01 $\pm <$0.01 dex kpc\textsuperscript{-1} for the 9.5 $< \log{\mathrm{M_{\star}}} <$ 10 and 10 $< \log{\mathrm{M_{\star}}} <$ 10.5 bins, respectively. For the highest mass bin, we are unable to measure a statistically robust profile with the imposed galaxy and spaxel cuts. Similar trends are found for a handful of galaxies at $z \sim$ 2 \citep{tacchella2015, tacchella2018}, where flat profiles\footnote{We note that profiles from \citet{tacchella2015, tacchella2018} are also measured in terms of sSFRs, different from the offset from the rSFMS used in this work and with work from MaNGA. While we do not perform the same sSFR comparison with their sample as we've done with \citet{nelson2016}, they are still comparable in that both measurements are normalised by stellar mass.} are observed on average for above MS galaxies but arising from a large variety for individual systems that appears to be mass dependent. Together, these results suggest that SF profiles of low to intermediate mass galaxies above the MS are flat out to $\sim2-3R_\mathrm{e}$ from cosmic noon until at least $\sim4$ Gyrs ago perhaps indicative of efficient gas accretion and transport throughout galaxies. While at $z\sim0$, galaxies above the MS may more regularly have centrally enhanced SF due to central starbursts partly driven by mergers. 

On the other hand, there is a strong agreement in the range of $\mathrm{\Delta \Sigma_{SFR}}$ values probed by the main sequence galaxies (blue) at $z\sim0.3$ and $z\sim0$. The two profiles are both flat within $R \leq 1.5~R_{\rm e}$, where the MaNGA and MAGPI profiles measure slopes of 0.00 $\pm$ 0.01 and 0.06 $\pm$ 0.01 dex $R_{\rm e}$\textsuperscript{-1}, respectively. While there is a statistically significant difference in the slopes based on the measured uncertainties, we note that the difference mainly stems from the profiles diverging from one another beyond $R \sim 1 \ R_{\rm e}$. This suggests that galaxies on the SFMS across the epochs probed are regulating their SF activity in a similar manner in the inner regions of galaxies. However, the uptick observed in the outskirts of the MAGPI sample may hint at external mechanisms, highlighting the importance of mapping SF activity out to large galactocentric radii. The positive slope in the MAGPI sample at large galactic radii is driven by a small number of galaxies (Section~\ref{sec:dsfrprof}) but is consistent with average sSFR profiles of massive galaxies at $z\sim1$ \citep{nelson2016} and individual profiles measured of massive galaxies at $z\sim2$ \citep{tacchella2018} which have also been measured to $\sim3 R_{\rm e}$. In particular, when converted to sSFRs, the MS (black) profiles measure flat slopes across all stellar mass bins (averaging $\approx$0.03 $\pm <$0.01 dex kpc\textsuperscript{-1}), which is in agreement with the profiles from \citet{nelson2016} (slopes averaging $<$0.1 dex kpc\textsuperscript{-1}).

The two lowest $\Delta$SFR bins (i.e., green and red profiles) show larger differences in $\mathrm{\Delta \Sigma_{SFR}}$ between $z = 0$ and $z = 0.3$. In general, we observe higher levels of star formation activity in our MAGPI sample for the same $\Delta$SFR bins.\footnote{Barring differences in the loci of the SFMS, which can rise from differences in methodology (\citet{bluck2020b} uses the definition from \citet{renzinipeng2015}, which does not require a pre-selection of SF galaxies as done in this study) and/or evolution of the SFMS over time, the bins are considered to be identical in ranges of $\Delta$SFR.} The offset in $\mathrm{\Delta \Sigma_{SFR}}$ could be due to differences in populations probed at the two epochs; recent results from the EAGLE simulation \citep{wang2023} show that galaxies with centrally concentrated SF below the SFMS are more common at $z = 0$ than at $z \approx$ 0.3. Centrally concentrated SF suggests relatively quenched outskirts of these galaxies due to environmental mechanisms, such that these findings suggest that there are more quenched satellites in the local Universe than at the MAGPI redshift regime. In terms of the measured slopes, there is good agreement between the MaNGA and MAGPI profiles \textit{within} $R \leq 1.5 \ R_{\rm e}$ for the -1.1 $< \Delta$SFR $<$ -0.5 bin (green), with measurements of 0.12 $\pm$ 0.01 and 0.08 $\pm$ 0.03 dex $R_{\rm e}$\textsuperscript{-1}, respectively. The MaNGA profile shows a stronger hint of inside-out quenching (i.e., steeper positive gradient), but we note that the MAGPI profile does measure a \textit{steeper} gradient (slope = 0.17 $\pm$ 0.02 dex $R_{\rm e}$\textsuperscript{-1}) \textit{out to} $R \sim \ 2 \ R_{\rm e}$. On the other hand, the MaNGA profile for the $\Delta$SFR $<$ -1.1 bin (red) is overall flat (slope = -0.02 $\pm <$0.01 dex $R_{\rm e}$\textsuperscript{-1}), whereas the MAGPI profile measures a slightly steeper negative gradient (slope = -0.09 $\pm$ 0.02 dex $R_{\rm e}$\textsuperscript{-1}) for $R \leq 1.5 \ R_{\rm e}$. Similar to the -1.1 $< \Delta$SFR $<$ -0.5 bin, the MAGPI profile shows a stronger hint of potential inside-out quenching with a measured slope of 0.11 $\pm$ 0.01 dex $R_{\rm e}$\textsuperscript{-1} when measured \textit{out to} 3 $R_{\rm e}$.

However, the caveat here is that the flatness of the $\Delta$SFR $<$ -1.1 profile from MaNGA largely rises from assigning a constant value to all D4000 $>$ 1.45 measurements. When a similar treatment is done on the -1.1 $< \Delta$SFR $<$ -0.5 and $\Delta$SFR $<$ -1.1 bins for the MAGPI sample, the profiles become much flatter and reside around $\mathrm{\Delta \Sigma_{SFR}} \sim$ -1.5. Given the considerable fraction of upper limits that exist in both bins, in particular for the $\Delta$SFR $<$ -1.1 bin (see Figure \ref{fig:upperlim}), the contrast in $\mathrm{\Delta \Sigma_{SFR}}$ between $z = 0$ and $z = 0.3$ does not necessarily indicate an evolution in SF. 

At $z\sim1-2$, a central (within $\sim1 R_\mathrm{e}$) depression is seen in sSFR profiles of galaxies below the SFMS particularly for the most massive galaxies (\citealt{nelson2016, tacchella2018}; see also the red dashed lines in the panels with $\log{\mathrm{M_{\star}}}>10$ in Figure \ref{fig:nelson16comp}). We observe mixed results with the sSFR profiles for MAGPI galaxies below the MS (red), in that the profiles in the 9 $< \log{\mathrm{M_{\star}}} <$ 9.5 and 10 $< \log{\mathrm{M_{\star}}} <$ 10.5 bins show positive gradients (slopes of 0.10 $\pm$ 0.02 and 0.08 $\pm$ 0.01 dex kpc\textsuperscript{-1}, respectively), whereas the profiles in the 9.5 $< \log{\mathrm{M_{\star}}} <$ 10 and 10.5 $< \log{\mathrm{M_{\star}}} <$ 11 bins show flatter profiles (slopes of -0.02 $\pm$ 0.01 and 0.01 $\pm <$0.01 dex kpc\textsuperscript{-1}). The lack of a trend with increasing stellar mass may come from a lack of robust statistics per radial bin, as apparent by the larger uncertainties and shorter radial distances probed. We also note that given the distribution of MAGPI galaxies below the SFMS in SFR-M\textsubscript{$\star$} space (see Figure \ref{fig:gsfms}), stellar mass bin widths of 0.5 dex in the range of 9 $< \log{\mathrm{M_{\star}}} <$ 11 naturally limit the number of available galaxies for analysis. However, when considered over the entire stellar mass range, galaxies at $z \sim$ 1 do show flat MS-normalised H$\alpha$ profiles, which is in line with the MAGPI and MaNGA results seen with the $\mathrm{\Delta \Sigma_{SFR}}$ profiles.

\begin{table*}
    \centering
    \caption{Table of measured slopes of $\mathrm{\Delta \Sigma_{SFR}}$ radial profiles for MaNGA and MAGPI}
    \begin{tabular}{cccc}
    \hline
     Global SF state & MaNGA slopes & MAGPI slopes ($R \leq$ 1.5 $R_{\rm e}$) & MAGPI slopes ($R \leq$ 3.0 $R_{\rm e}$) \\
     (dex) & (dex $R_{\rm e}$\textsuperscript{-1}) & (dex $R_{\rm e}$\textsuperscript{-1}) & (dex $R_{\rm e}$\textsuperscript{-1}) \\ \hline
     $\Delta$SFR $>$ 0.5 & -0.34 $\pm$ 0.02 & -0.02 $\pm$ 0.02 & -0.05 $\pm$ 0.01 \\
     -0.5 $<$ $\Delta$SFR $<$ 0.5 & 0.00 $\pm$ 0.01 & 0.06 $\pm$ 0.01 & 0.11 $\pm$ $<$0.01 \\
     -1.1 $<$ $\Delta$SFR $<$ -0.5 & 0.12 $\pm$ 0.01 & 0.08 $\pm$ 0.03 & 0.17 $\pm$ 0.02 \\
     $\Delta$SFR $<$ -1.1 & -0.02 $\pm$ $<$0.01 & -0.09 $\pm$ 0.02 & 0.11 $\pm$ 0.01 \\
    \hline
    \end{tabular}
    \label{tab:mangaslopes}
\end{table*}

\subsection{What regulates and/or quenches star formation at \texorpdfstring{$z \boldsymbol{\sim}$}{sim} 0.3?}
\label{sec:quench}

We observe enhanced star formation at all radial distances probed with respect to the resolved SFMS in the most star-forming bin (i.e., $\Delta$SFR $>$ 0.5). Given that all galaxies contributing to the profile have H$\alpha$-measured $\mathrm{\Sigma_{SFR}}$, this suggests that MAGPI galaxies in this bin have sufficient cold gas reservoirs to fuel star formation as traced at this epoch. The overall flatness (slope = -0.05 $\pm$ 0.01 dex $R_{\rm e}$\textsuperscript{-1}) of the profile suggests that the physical processes at play for these galaxies do not lead to a preferred radial trend in star formation enhancement. 

Previous studies (\citealt{mh1996, moreno2015}) have proposed that gravitational torques can transport the newly accreted gas from these interactions towards the centre, eventually leading to a starburst. This can be caused partly by galaxy-galaxy and/or tidal interactions, triggering inflow of cold gas external to the galaxy. To test this hypothesis, we search for nearby neighbours in both redshift and on-sky separations. Following the findings of \citet{patton2013}, we first search for nearby galaxies with line-of-sight velocities of $\Delta v$ $<$ 1000 km s\textsuperscript{-1} and projected distances of $\lesssim$150 kpc $-$ the maximum projected separation out to which \citet{patton2013} found statistically significant SF enhancements. This simple test indicates that $\sim$78 per cent (21/27) of galaxies in the $\Delta$SFR $>$ 0.5 bin have at least one nearby neighbour, some having as many as 11 potential neighbours. We also confirm these results with a preliminary group finding analysis, where we find $\sim$85 per cent (23/27) of galaxies belong to groups ranging between 2 to 12 members. If galaxy-galaxy interactions are the dominant process at play, the median profile for this bin would show a negative gradient due to the central starburst. Given the flatness of the profile, this suggests that interactions with nearby neighbours cannot be the sole physical process at work for star-forming galaxies residing above the SFMS. 

In contrast, a generally negative trend in star formation activity with increasing galactocentric radii for galaxies above the SFMS has been observed in the local Universe (i.e., $z \sim$ 0). In particular, the central enhancement of star formation activity in MaNGA galaxies can be associated with metal-poor gas accretion \citep{ellison2018}. This suggests that major and minor mergers, along with interactions with dwarf satellites, are some of the likely causes of the inflow of pristine gas towards the centre. The central enhancement is confirmed by \citet{wang2019}, where they find shorter gas depletion times centrally compared to the outskirts. Assuming a simple gas regulator model, these results suggest that spatial variation in SFRs in galaxies may simply be due to variation in SF efficiencies, such that any changes in gas accretion rates are immediately reflected upon the measured SFRs. On the other hand, \citet{nelson2016} do not find evidence for strong radial trends in star formation as probed by EW(H$\alpha$) at $z \sim$ 1. They instead find that star formation is enhanced throughout the galaxy disc for galaxies located above the main sequence, which is suggestive of physical processes not exclusive to major mergers and galaxy interactions. We find our results to be more consistent with those of \citet{nelson2016}, which suggests that radial trends already start to evolve at the MAGPI redshift regime. This could also be explained by the fact that gas accretion rates are overall higher at higher redshifts \citep[e.g.,][]{vandevoort2011}. 

Galaxies on the main sequence (i.e., -0.5 $<$ $\Delta$SFR $<$ 0.5) show a positive gradient in $\mathrm{\Delta \Sigma_{SFR}}$ with a slope of 0.11 $\pm <$0.01 dex $R_{\rm e}$\textsuperscript{-1}. Generally small bootstrap uncertainties of the median profile suggest that local variation in $\mathrm{\Sigma_{SFR}}$ is minimal and that values are generally distributed around $\mathrm{\Delta \Sigma_{SFR}}$ = 0. Relatively small uncertainties in the luminosity-weighted age profile also suggest that the increase in age with decreasing radius towards the centre is likely real. These results are consistent with an inside-out growth scenario, where galaxies start building up their stellar mass in the centre. 

A model based on high-redshift ($z \sim$ 2 - 4) simulations suggests that galaxies on the SFMS undergo multiple cycles of gas inflow and depletion in the centre, which results in oscillations about the SFMS and as such scatter about the main sequence \citep{tacchella2016}. It is not until galaxies reach a critical virial mass of $\sim$10\textsuperscript{11.5} M\textsubscript{\sun}, where galaxies then begin a full quenching process triggered by shock heating of the halo. On the other hand, analysis based on the EAGLE simulations \citep{mattheeschaye2019} at $z \sim 0.1$ suggests that the scatter of the SFMS is mostly modulated by long timescale ($\sim$10 Gyr) variations that depend on the halo's assembly history. Fluctuations due to self-regulation processes such as cooling, SF, and outflows run on shorter timescales and are only important for less massive galaxies. Both scenarios apply to our sample, in that any competing physical mechanisms at play likely cancel one another out and that any long timescale processes will not be reflected in H$\alpha$-based measurements, given that H$\alpha$ traces timescales as short as $\sim$10 Myr. With the former, this implies a balance between gas accretion and depletion due to star formation and/or galactic winds, which is unsurprising for galaxies located on the SFMS. Furthermore, our nearby neighbour search suggests that 65 per cent (85/131) of galaxies in this $\Delta$SFR bin have neighbours within projected separations of 150 kpc or lower, giving further support for external cold gas accretion. Results from the group finding analysis give similar numbers, where some groups consist of as many as 20 members. 

Similar conclusions were reached in another study at a similar redshift range of 0.1 $< z <$ 0.42 with the MUSE-Wide survey, which show evidence for inside-out growth in star-forming galaxies based on observing a positive gradient in dust-corrected H$\alpha$-based sSFRs \citep{jafariyazani2019}. Furthermore, observations of extended H$\alpha$ emission compared to the stellar continuum, along with increasingly centrally depressed H$\alpha$ equivalent widths with stellar mass, at $z \sim$ 1 suggests that inside-out assembly of galaxy discs are ubiquitous across a broad range of redshifts \citep{nelson2016, matharu2022}.

While galaxies in the higher star-forming bins show evidence of star formation fueled by consistent cold gas accretion, galaxies in the -1.1 $< \Delta$SFR $<$ -0.5 bin begin to show signs of quenching across all radial bins probed. The positive gradient in $\mathrm{\Delta \Sigma_{SFR}}$ suggests inside-out quenching, with relatively higher SF measured in the outskirts. This trend suggests that it is moreso internal rather than external mechanisms dominating the SF activity in these galaxies. As mentioned previously in Section \ref{sec:data}, the MAGPI sample selected for this study do not include galaxies in the most dense environments, i.e., galaxy clusters. Majority of our galaxies ($\sim$76 per cent) belong to groups, where environmental mechanisms such as ram pressure stripping \citep{gunngott1972, kk2004a, kk2004b, bg2006}, viscous stripping \citep{nulsen1982}, and starvation/strangulation \citep{larson1980, bekki2002}, are not expected to be as strong as they are in cluster environments. Gas stripping mechanisms in particular are expected to be more efficient in the galaxy outskirts, where the gas is less tightly confined to the galaxy's gravitational potential. The aftermath of such mechanisms would manifest itself in a negative gradient in SF, characteristic of quenching starting from outside-in. For this $\Delta$SFR bin in particular, $\sim$69 per cent (29/42) of them belong to groups, some with as many as 20 members. Given that we do not observe a negative gradient, we attribute the overall low levels of SF and the positive gradient in $\mathrm{\Delta \Sigma_{SFR}}$, to `pre-processing' \citep{fujita2004} -- where galaxies are subject to interactions with other galaxies and the group potential itself, leading to SF quenching. 

Clear features of environmental mechanisms at play are observed in the local Universe. For MaNGA, notable differences in $\mathrm{\Delta \Sigma_{SFR}}$ profiles between the centrals and satellites for green valley (i.e., galaxies in the -1.1 $<$ $\Delta$SFR $<$ -0.5 bin) galaxies have been observed \citep{bluck2020b}. While both centrals and satellites exhibit quiescent cores, satellites do not have star-forming outskirts as centrals do. Nevertheless, satellites in the MaNGA sample do exhibit positive $\mathrm{\Delta \Sigma_{SFR}}$ gradients (slope = 0.12 $\pm$ 0.01 dex $R_{\rm e}$\textsuperscript{-1}; Table \ref{tab:mangaslopes}) overall. Evidence for outside-in quenching is only captured in low-mass (i.e., M\textsubscript{$\star$} $<$ 10\textsuperscript{10} M\textsubscript{\sun}) satellites for the MaNGA sample. These results imply that environmental processes play a more dominant role for low-mass satellites, which may explain for the lack of a clear outside-in quenching feature in the MAGPI sample. Due to current sample statistics in both galaxies and spaxels, we are unable to perform a similar analysis for the current MAGPI sample. 

We also mentioned previously that the entire $\mathrm{\Delta \Sigma_{SFR}}$ profile for this bin is considered an upper limit, given the limitation of the D4000-sSFR relation. In Appendix \ref{sec:app1}, we show that the majority of spaxels with upper limits are distributed around $\mathrm{\Delta \Sigma_{SFR}}$ $\simeq$ -1.5, with the largest fraction being in the 0.5 -- 1.0 $R_{\rm e}$ bin. In terms of the ratio of detections vs. non-detections, the contribution from both are comparable such that the entire profile will likely be shifted to lower $\mathrm{\Delta \Sigma_{SFR}}$ with proper consideration of upper limits. We do not expect to see a different trend other than the previously observed positive gradient (i.e., inside-out quenching). We also address the effects of combining H$\alpha$- and D4000-based $\mathrm{\Sigma_{SFR}}$ measurements on the $\mathrm{\Delta \Sigma_{SFR}}$ profile in Appendix \ref{sec:app1}. Our comparison of H$\alpha$- and D4000-only profiles show that while D4000 alone does systematically underestimate the level of SF, there is in general a good agreement between the two measurements (Figures \ref{fig:gvcomp} and \ref{fig:gvsamecomp}). 

Finally, we discuss the implications of the galaxies in the least star-forming bin, i.e., $\Delta$SFR $<$ -1.1 (red). The measured slope of 0.11 $\pm$ 0.01 dex $R_{\rm e}$\textsuperscript{-1} of the $\mathrm{\Delta \Sigma_{SFR}}$ profile suggests that the central region is relatively quenched in comparison to the outskirts. The profile suggests physical processes that are preferentially at work within the galaxy centre, which may include AGN feedback and/or expedited use of star-forming gas. While we have removed any spaxels classified as optical AGN, it is possible that we may have missed out on weaker AGN emission due to lack of sufficient S/N (i.e., S/N $\geq$ 3) in one or more emission lines, or AGN identifiable by multi-wavelength tracers. The lack of a current optical AGN signature also does not preclude previous activity of an AGN that is currently not active \citep{padovani2017}. 

In local galaxies, $\mathrm{\Delta \Sigma_{SFR}}$ profiles in the most and least star-forming bins mirror one another in that there is a clear central enhancement in the former and suppression in the latter \citep{ellison2018}. This has been interpreted as an indication of inside-out mass growth triggered by gas inflow preferentially towards the central regions, followed by eventual quenching propagating from inside-out. A variety of scenarios could explain this phenomenon, which include gas depletion and AGN feedback. On the other hand, \citet{wang2019} associate central enhancement/suppression in galaxies above/below the SFMS with different response rates to changes in gas accretion rates from the halo; positive/negative SF gradients are thus not necessarily hinting towards different physical processes at play. However with the same parent dataset, \citet{bluck2020b} observe a much flatter profile for galaxies of the same bin (i.e., their $\Delta$SFR $<$ -1.1 bin), where the lack of a clear trend seem to come from their choice of extending the D4000-sSFR relation to higher D4000 (i.e., D4000 $>$ 1.45) by assigning the same artificial value of sSFR = 10\textsuperscript{-12} yr\textsuperscript{-1} \citep[see][Figure 4]{bluck2020a}. We can see this effect at work within our own sample. When we include only the spaxels with D4000 $\leq$ 1.475 where the relation can be defined, we see the central depression. However, when we include spaxels with a measured D4000 beyond this limit as upper limits defined by a floor in sSFR at high D4000, we obtain a flat profile consistent with \citet{bluck2020b}. As shown in Appendix \ref{sec:app1}, we observe a higher fraction of upper limits towards the galaxy centre (i.e., 0 -- 1.5 $R_{\rm e}$). This suggests that with improved measurements on the upper limits, the $\mathrm{\Delta \Sigma_{SFR}}$ profile is likely to show a \textit{steeper} positive gradient consistent with inside-out quenching. 

\section{Summary}
\label{sec:summ}
In this study, we have investigated the radial trends in $\mathrm{\Delta \Sigma_{SFR}}$ and Age\textsubscript{L} as a function of location with respect to the SFMS (i.e., $\Delta$SFR) to constrain the physical processes enhancing, regulating, and quenching MAGPI galaxies. We use D4000 in addition to the H$\alpha$ emission line to ensure that we capture low levels of SF, such that our sample is as complete as possible in the ranges of global SF states probed. We also place our results in context with galaxies at low- ($z \sim$ 0) and high-redshift ($z \sim$ 1 and $z \sim$ 2) to investigate the possibility of an evolution in radial trends in SF activity across these redshift regimes. We summarise our key results below: 
\begin{enumerate}
    \item Star-forming galaxies above the SFMS (i.e., $\Delta$SFR $>$ 0.5) at the MAGPI redshift regime show uniformly enhanced SF throughout all radial distances probed, suggesting a mixture of physical processes (e.g., galaxy-galaxy interactions) at play along with a sufficient cold gas reservoir to fuel SF. 
    \item Galaxies along the SFMS (i.e., -0.5 $<$ $\Delta$SFR $<$ 0.5) show a slightly positive gradient in $\mathrm{\Delta \Sigma_{SFR}}$, where the uptick towards the outskirts is associated with a few highly star-forming galaxies. Otherwise, the inner regime of the profile is flat as expected from MS galaxies maintaining their SF via both a balance between gas accretion and depletion \citep{tacchella2016} and long timescale variations due to differences in halo assembly histories \citep{mattheeschaye2019}. The negative gradient in luminosity-weighted stellar ages (i.e., Age\textsubscript{L}) also suggests inside-out growth \citep{munoz-mateos07, vandokkum2010}, in agreement with other studies \citep{nelson2016, jafariyazani2019, matharu2022} as well. 
    \item Galaxies just below the SFMS (i.e., -1.1 $< \Delta$SFR $<$ -0.5) show a positive gradient in SF, suggestive of inside-out quenching. Our results suggest that galaxies likely have undergone pre-processing from interactions with other galaxies and the group gravitational potential. We also compare H$\alpha$ and D4000-only profiles to find that D4000-measured $\mathrm{\Delta \Sigma_{SFR}}$ profiles are systematically lower but do not completely dominate in spaxel contribution in any specific radial bin. 
    We attribute the lack of outside-in quenching feature to the majority of our sample galaxies being in the central regions of group environments.
    \item Quenched galaxies (i.e., $\Delta$SFR $<$ -1.1) in the MAGPI sample show a positive gradient in $\mathrm{\Delta \Sigma_{SFR}}$ as well, where the central dip in SF suggests the aftermath of AGN feedback or expedited use of star-forming gas due to a nuclear starburst. We also consider the possibility of simply lower SF efficiencies leading to lower levels of SF in the centre \citep{wang2019}. Given that the majority of galaxies in this bin have $\mathrm{\Delta \Sigma_{SFR}}$ measured from D4000, we acknowledge that we are missing a large fraction of the quenched population due to limitations in the measured D4000-sSFR relation (see Section \ref{sec:quench}, but also \citealt{bluck2020a, thorp2022}). With proper consideration of upper limits, we expect to see stronger hints of inside-out quenching for quenched MAGPI galaxies. 
    \item We capture a potential evolution in the radial trends in SF for star-forming galaxies from $z \sim$ 2 to $z \sim$ 0. The flat $\mathrm{\Delta \Sigma_{SFR}}$ observed in star-forming MAGPI galaxies is in contrast to the results observed in the local Universe \citep{ellison2018, bluck2020b}, where galaxies clearly show centrally enhanced SF and thus a negative gradient in SF. Agreements with observations at $z \sim 1-2$ \citep{nelson2016, tacchella2015, tacchella2018} suggest that this may be an indicator of an evolution in the radial trends in SF over the last 4 Gyrs.
\end{enumerate}

In future work, we plan to place constraints on the physical mechanisms governing MAGPI galaxies by exploring matched samples within a suite of cosmological simulations (Harborne et al. in prep) and to complete a more statistical environmental analysis using environmental metrics probed out to 5 Mpc (Barsanti et al. in prep). 

\section*{Acknowledgements}
We would like to thank the anonymous referee for their helpful comments that improved the manuscript. MM thanks Mallory D. Thorp for helpful discussions on the use of D4000 as a SFR indicator. MM thanks Asa F. L. Bluck for sharing his MaNGA radial profiles as a comparison set with the MAGPI results. MM also extends thanks to Erica J. Nelson for approving the use of her 3D-\textit{HST} profiles as a comparison set. We wish to thank the ESO staff, and in particular the staff at Paranal Observatory, for carrying out the MAGPI observations. MAGPI targets were selected from GAMA. GAMA is a joint European-Australasian project based around a spectroscopic campaign using the Anglo-Australian Telescope. GAMA was funded by the STFC (UK), the ARC (Australia), the AAO, and the participating institutions. GAMA photometry is based on observations made with ESO Telescopes at the La Silla Paranal Observatory under programme ID 179.A-2004, ID 177.A-3016. The MAGPI team acknowledge support by the Australian Research Council Centre of Excellence for All Sky Astrophysics in 3 Dimensions (ASTRO 3D), through project number CE170100013. CL, JTM and CF are the recipients of ARC Discovery Project DP210101945. CF is the recipient of an Australian Research Council Future Fellowship (project number FT210100168) funded by the Australian Government. AG would like to acknowledge support by the Australian Research Council Centre of Excellence for All Sky Astrophysics in 3 Dimensions (ASTRO 3D), through project number CE170100013. LMV acknowledges support by the German Academic Scholarship Foundation (Studienstiftung des deutschen Volkes) and the Marianne-Plehn-Program of the Elite Network of Bavaria. KG is supported by the Australian Research Council through the Discovery Early Career Researcher Award (DECRA) Fellowship (project number DE220100766) funded by the Australian Government. PS is supported by the Leiden University Oort Fellowship and the International Astronomical Union TGF Fellowship. SMS acknowledges funding from the Australian Research Council (DE220100003).

This work made use of \textsc{Astropy}:\footnote{\href{http://www.astropy.org}{http://www.astropy.org}} a community-developed core Python package and an ecosystem of tools and resources for astronomy \citep{astropy2013, astropy2018, astropy2022}. This research has made use of the NASA/IPAC Infrared Science Archive \citep{https://doi.org/10.26131/irsa537}, which is funded by the National Aeronautics and Space Administration and operated by the California Institute of Technology. Parts of the results in this work make use of the colormaps in the \textsc{CMasher} \citep{vandervelden2020} package.

\section*{Data Availability}
All MUSE data present in this work are publicly available on the ESO archive. Data products such as fully reduced datacubes and emission line fits will be made available as part of a MAGPI team data release (Mendel et al. in prep; Battisti et al. in prep). 



\bibliographystyle{mnras}
\bibliography{biblio} 




\appendix

\section{Using D4000 as a SFR indicator}
\label{sec:app1}

\subsection{Reliability of D4000 derived SFRs}
\begin{figure}
    \centering
    \includegraphics[width=\columnwidth]{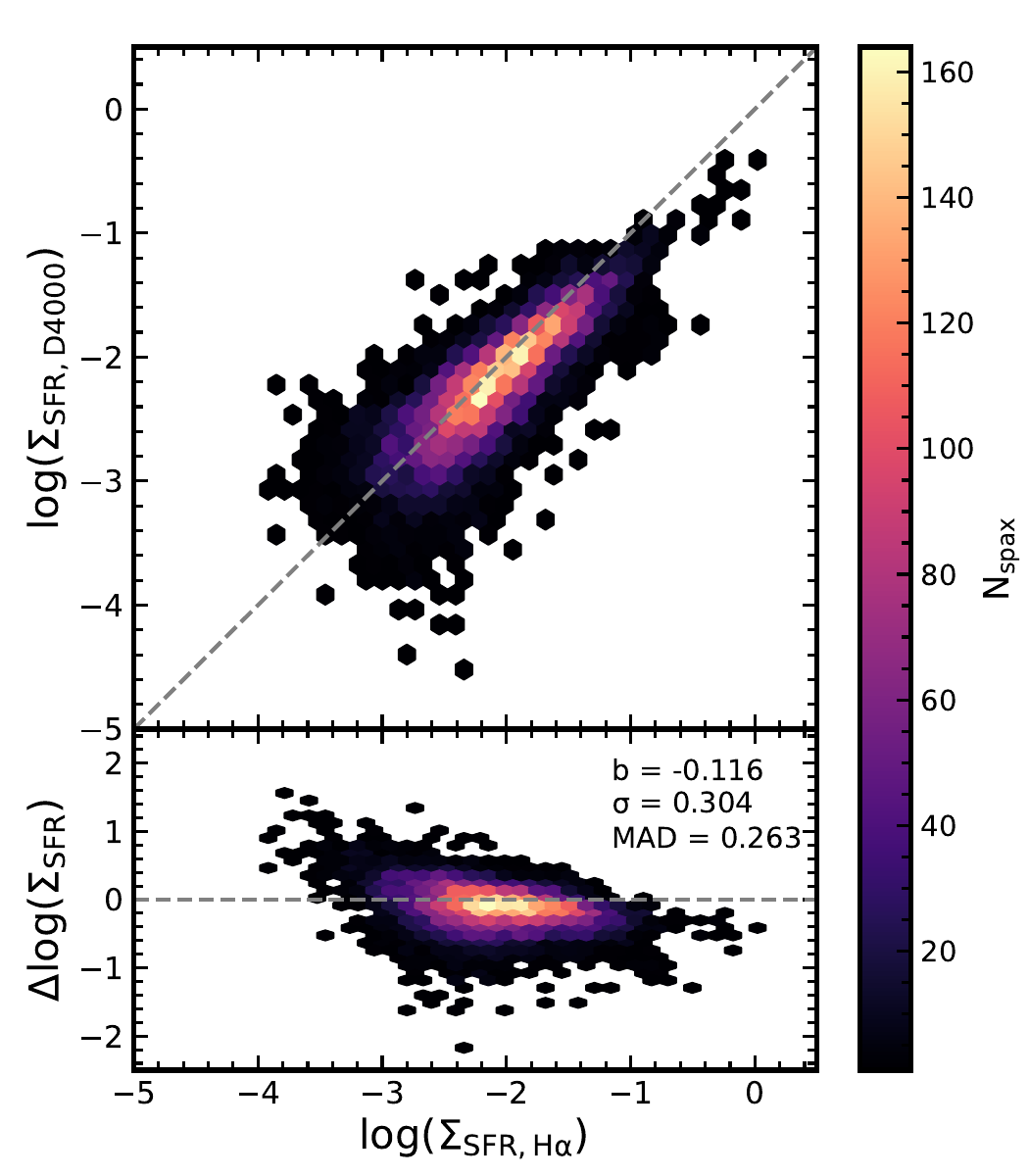}
    \caption{Comparison diagram of $\mathrm{\Sigma_{SFR, D4000}}$ and $\mathrm{\Sigma_{SFR, H\alpha}}$ for star-forming spaxels. The hexagonal bins are colour-coded by the number of spaxels. The grey dashed line indicates the 1:1 line. The bottom panel shows the distribution of $\mathrm{\Delta \log}$($\mathrm{\Sigma_{SFR}}$) = $\mathrm{\log}$($\mathrm{\Sigma_{SFR, D4000}}$) - $\mathrm{\log}$($\mathrm{\Sigma_{SFR, H\alpha}}$). The bias of $\mathrm{\Delta \log}$($\mathrm{\Sigma_{SFR}}$) is -0.116 dex, whereas we present the degree of scatter in standard deviation and MAD standard deviation, which are 0.304 and 0.263 dex, respectively, as shown in the same panel.}
    \label{fig:d4hacomp}
\end{figure}

We investigate the reliability of D4000 as a SFR indicator to ensure that the use of D4000 is not introducing significant bias in the analysis. In Figure \ref{fig:d4hacomp}, we show a plot comparing the $\mathrm{\Sigma_{SFR, H\alpha}}$ and $\mathrm{\Sigma_{SFR, D4000}}$ for all spaxels with reliably measured $\mathrm{\Sigma_{SFR, H\alpha}}$ in our sample. With the exception of some scatter as shown by the black hexagonal bins, the general trend is that both measurements agree well with one another with a bias of approximately -0.116 dex. We also present the scatter in standard deviation and the MAD standard deviation in the bottom panel, where both measurements are generally in good agreement as well. We use the standard deviation of the scatter to take into consideration the uncertainty rising from the recovery of H$\alpha$-based SFRs from D4000, by adding this value in quadrature when measuring errors on D4000-based SFRs, as explained in Section \ref{sec:d4}. 

\begin{figure}
    \centering
    \includegraphics[width=\columnwidth]{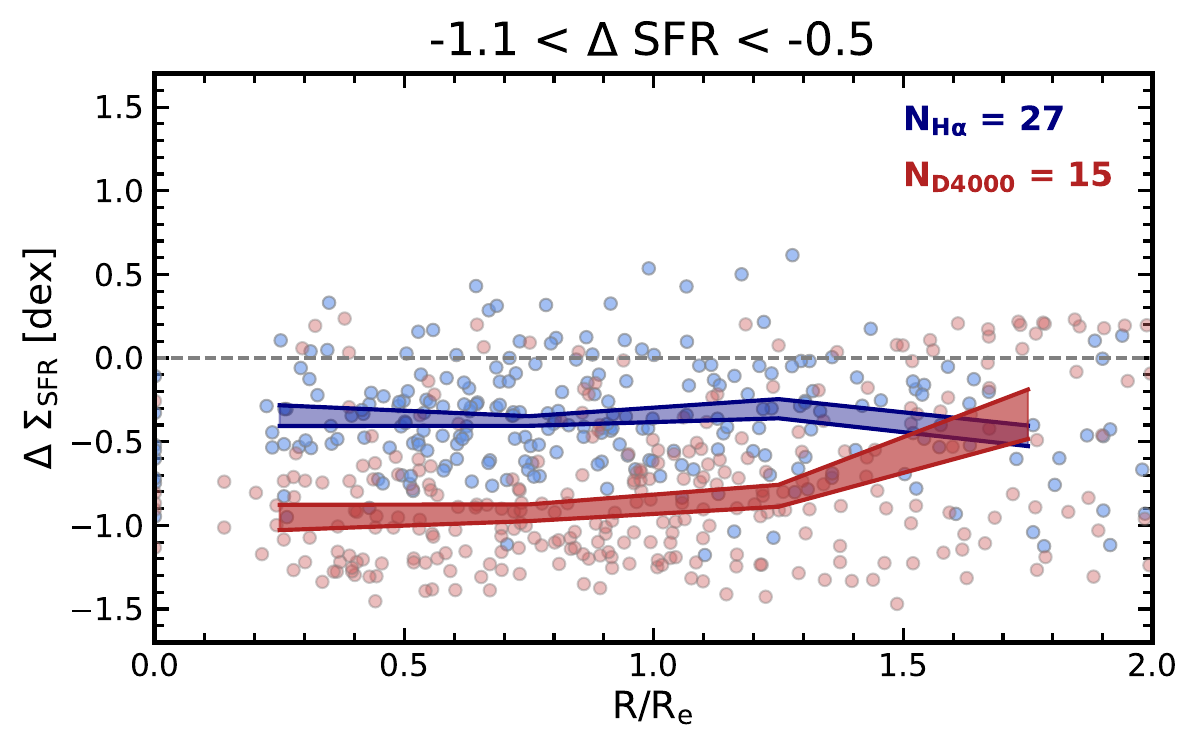}
    \caption{Median $\mathrm{\Delta \Sigma_{SFR}}$ radial profiles of galaxies in the -1.1 $<$ $\Delta$SFR $<$ -0.5 bin, where the blue points and line show the profile measured solely from H$\alpha$-derived $\mathrm{\Sigma_{SFR}}$ (i.e., H$\alpha$-detected and identified as SF by the BPT diagram; total of 27 galaxies) and the red points and line show the profile from D4000-derived $\mathrm{\Sigma_{SFR}}$ (i.e., either non-H$\alpha$-detected or H$\alpha$-detected but not classified as SF; total of 15 galaxies). Shaded regions indicate the range of bootstrap errors of the median profiles. The comparison of the two profiles show that the use of D4000 alone systematically underestimates the level of star formation activity, in comparison to H$\alpha$, which likely arises from different timescales probed by the two SFR indicators.}
    \label{fig:gvcomp}
\end{figure}

An important caveat to the $\mathrm{\Delta \Sigma_{SFR}}$ profile in the -1.1 $<$ $\Delta$SFR $<$ -0.5 (green) bin is that we use a combination of H$\alpha$- and D4000-based measurements. While the two have been shown to be well calibrated to each other (see Figure \ref{fig:d4hacomp}), it is possible that combining the different indicators may affect the radial profile. In total, $\sim$45 per cent of spaxels from galaxies in this bin consist of H$\alpha$-measured SFRs. In contrast, the $\Delta$SFR $>$ 0.5 (pink) bin and the -0.5 $<$ $\Delta$SFR $<$ 0.5 (blue) bin are dominated by spaxels with SFRs measured by H$\alpha$, 100 per cent and 99 per cent, respectively. While the $\Delta$SFR $<$ -1.1 bin (red) is dominated by spaxels with SFRs measured by D4000, $\sim$92 per cent. In Figure \ref{fig:gvcomp}, we split the -1.1 $<$ $\Delta$SFR $<$ -0.5 bin into H$\alpha$- (N = 27) and D4000-measured (N = 15) galaxies to investigate the implications of the resulting median $\mathrm{\Delta \Sigma_{SFR}}$ profiles. The D4000-based profile is systematically lower than that of H$\alpha$, as expected. However, both profiles span the full radial range of the -1.1 $<$ $\Delta$SFR $<$ -0.5 bin, suggesting that neither SFR indicator is biased towards either the centre or the outskirts. 

\begin{figure}
    \centering
    \includegraphics[width=\columnwidth]{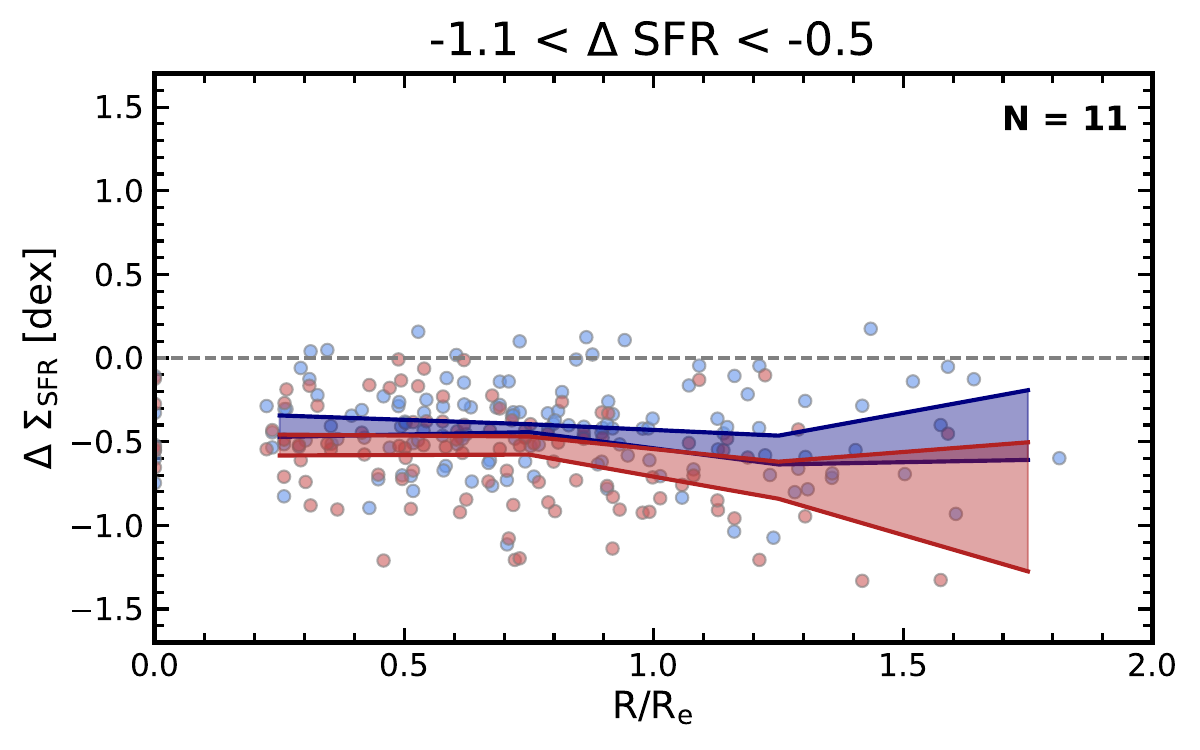}
    \caption{Median $\mathrm{\Delta \Sigma_{SFR}}$ radial profiles of 11 H$\alpha$-detected galaxies (i.e., subsample from the total of 27 galaxies shown in Figure \ref{fig:gvcomp}) with measured D4000 in the -1.1 $<$ $\Delta$SFR $<$ -0.5 bin. Using the same colour scheme as Figure \ref{fig:gvcomp}, the blue shaded region shows the median profile measured from H$\alpha$, and the red shaded region shows the profile measured from D4000. Overall, there is generally good agreement in the range of $\mathrm{\Delta \Sigma_{SFR}}$ values probed by both SFR indicators near the galaxy centre. The profiles show to diverge from one another towards the outskirts, where D4000-based measurements decline and H$\alpha$-based increase with increasing galactocentric radius. However, the profiles do overlap with one another throughout all radii probed, suggesting that the range of $\mathrm{\Delta \Sigma_{SFR}}$ values are generally in good agreement.}
    \label{fig:gvsamecomp}
\end{figure}

We re-calculate the $\mathrm{\Delta \Sigma_{SFR}}$ median profiles for 11 galaxies with measurements in both SFR tracers, as shown in Figure \ref{fig:gvsamecomp}. There is in general good agreement in the range of $\mathrm{\Delta \Sigma_{SFR}}$ measured by the two tracers for the same galaxies, although H$\alpha$-based $\mathrm{\Sigma_{SFR}}$ measurements do indicate slightly higher SF activity. Moreover, while the D4000 profile does show a slightly decreasing trend with radial distance, the number of spaxels with measured $\mathrm{\Sigma_{SFR}}$ also tapers off towards the outskirts, such that the profiles do agree within the range of uncertainties. 

\subsection{Radial distribution of SFR indicators}
\begin{figure}
    \centering
    \begin{subfigure}{\columnwidth}
        \centering
        \includegraphics[width=\columnwidth]{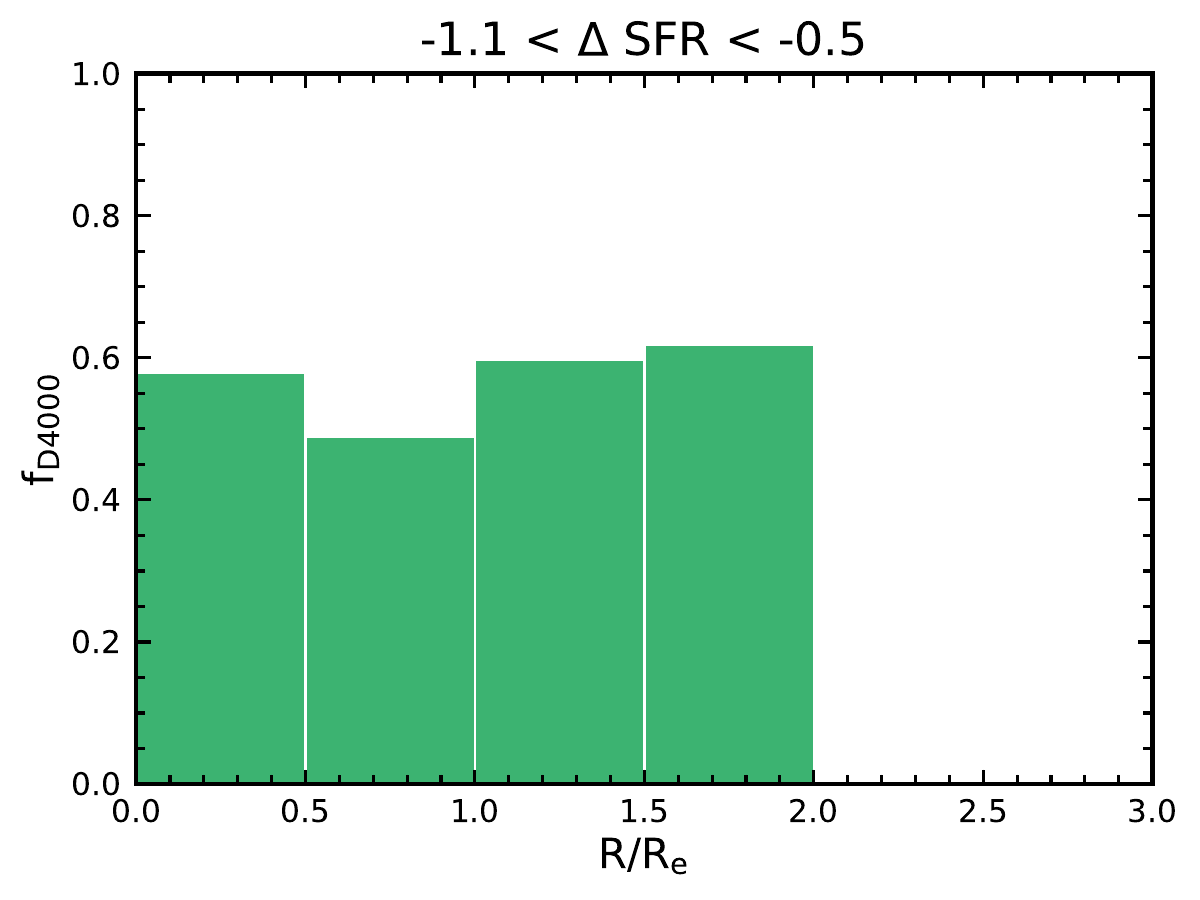}
    \end{subfigure}
    \hfill
    \begin{subfigure}{\columnwidth}
        \centering
        \includegraphics[width=\columnwidth]{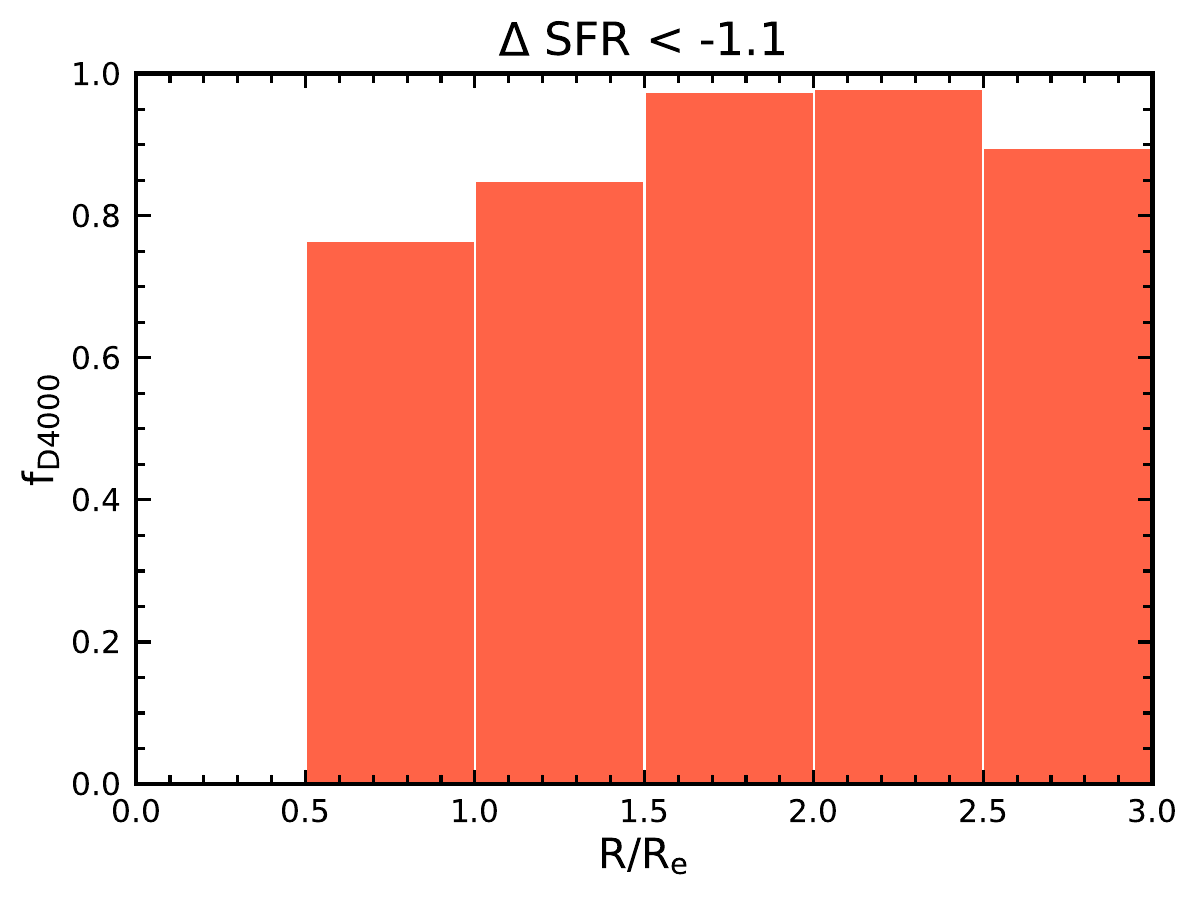}
    \end{subfigure}
    \caption{Contribution of D4000 spaxels per radial bin for the -1.1 $< \Delta$SFR $<$ -0.5 bin (top) and $\Delta$SFR $<$ -1.1 bin (bottom).}
    \label{fig:d4frachist}
\end{figure}

We explore whether D4000 measurements are preferentially located in certain radial regimes of our sample galaxies for the lower star-forming bins of -1.1 $< \Delta$SFR $<$ -0.5 and $\Delta$SFR $<$ -1.1, where the contribution of D4000 measurements notably increase. We investigate the fraction of D4000-measured $\mathrm{\Sigma_{SFR}}$ (i.e., f\textsubscript{D4000}) as a function of galactocentric radii ($R$/$R_{\rm e}$) and show the resulting histograms in Figure \ref{fig:d4frachist}, where the top panel is for the -1.1 $< \Delta$SFR $<$ -0.5 (green) bin and the bottom for the $\Delta$SFR $<$ -1.1 (red) bin. For the -1.1 $< \Delta$SFR $<$ -0.5 bin, we find that f\textsubscript{D4000} is uniform across all radial bins, suggesting that D4000 measurements are not introducing any biases in the $\mathrm{\Delta \Sigma_{SFR}}$ profile. For the $\Delta$SFR $<$ -1.1 bin, f\textsubscript{D4000} is notably higher across all radial bins, with relatively higher fractions observed in the 1.5 -- 2.5 $R_{\rm e}$ regimes. However, the fractional contribution of D4000 measurements is still comparable among all radial bins, such that we reach the same conclusion that the use of D4000 as an SFR indicator is not biasing our measurements to particular spatial regimes within our galaxy sample. 

\subsection{Upper limits on D4000-based SFRs}

\begin{figure*}
    \centering
    \begin{subfigure}{\columnwidth}
        \centering
        \includegraphics[width=\columnwidth]{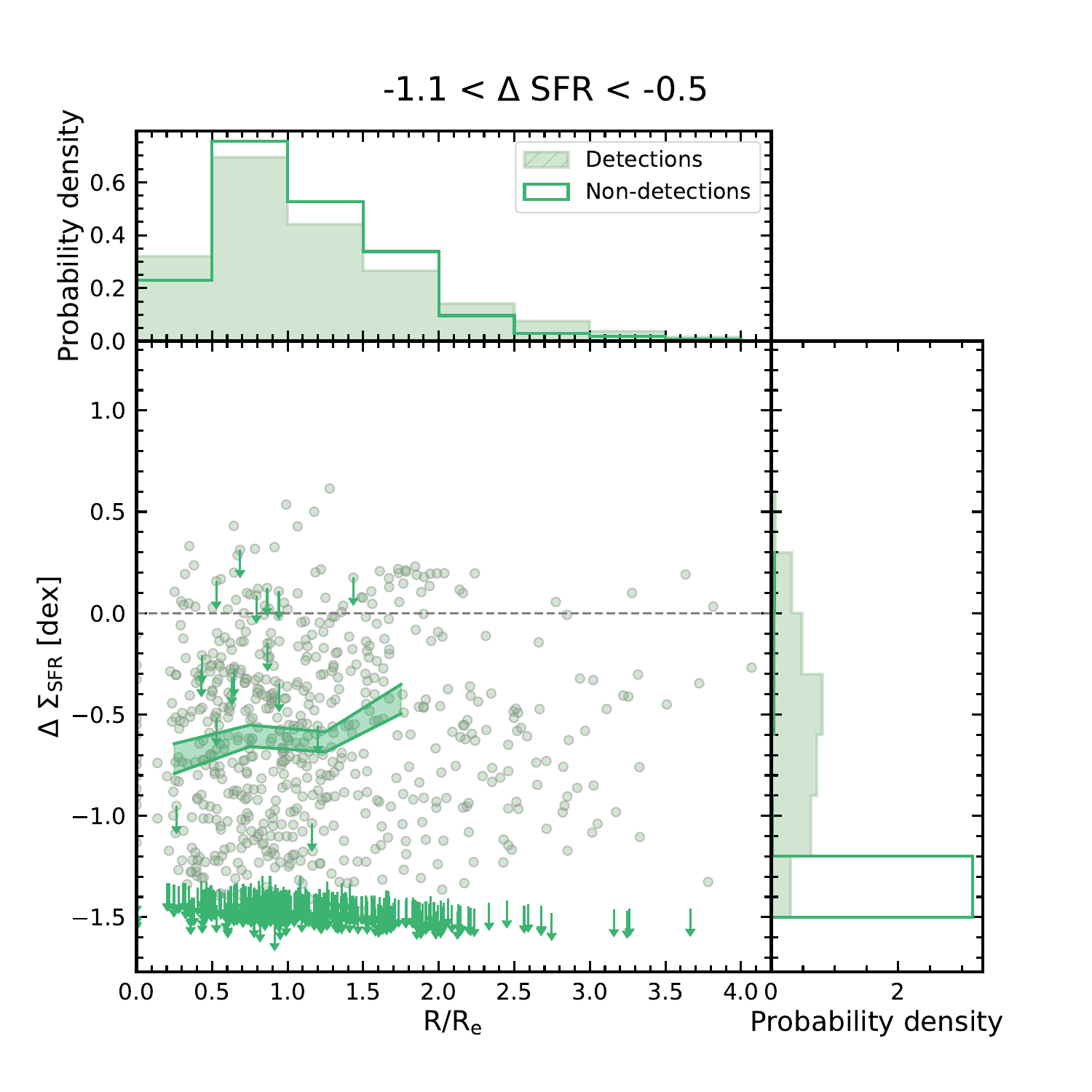}
    \end{subfigure}
    \hfill
    \begin{subfigure}{\columnwidth}
        \centering
        \includegraphics[width=\columnwidth]{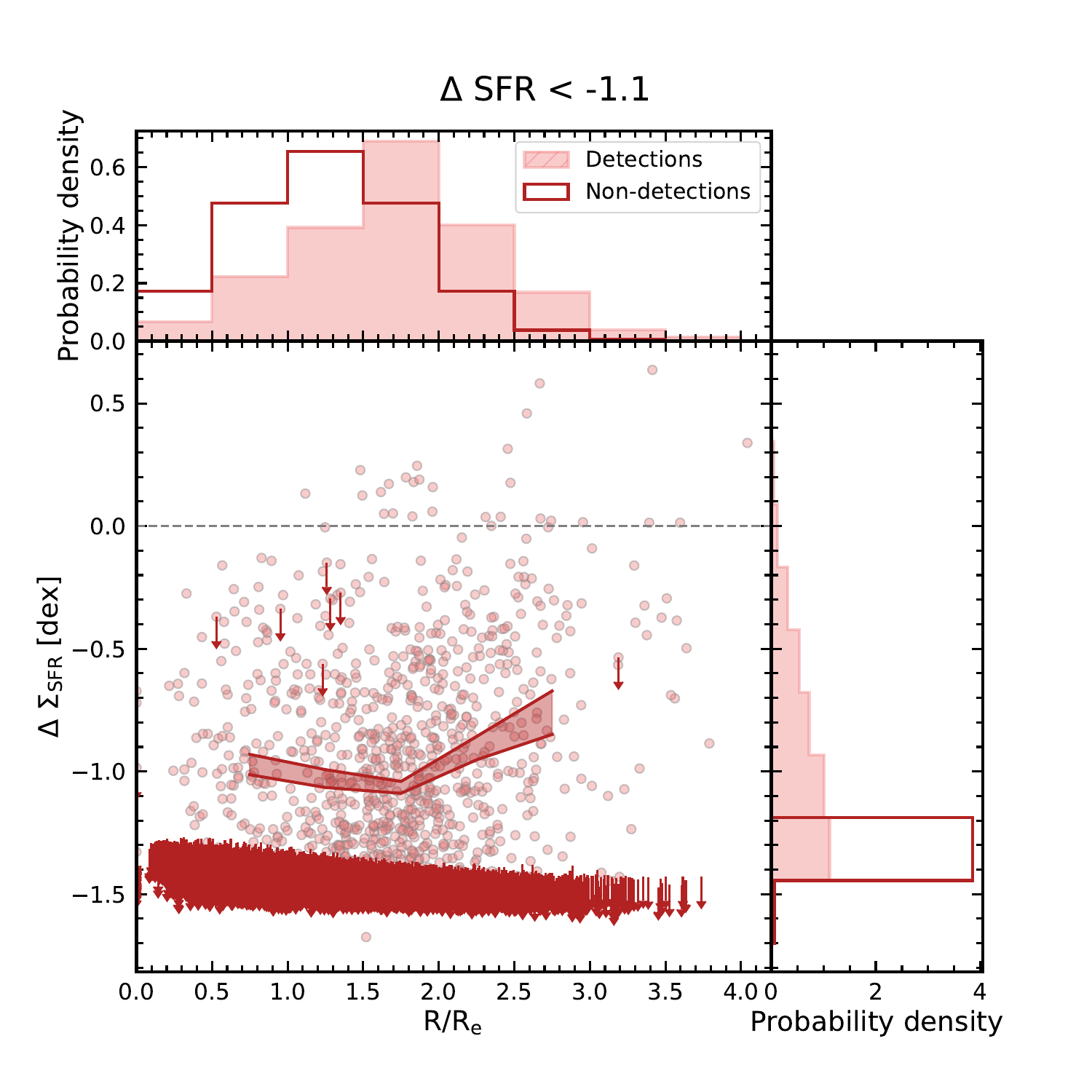}
    \end{subfigure}
    \caption{Distribution of detections and non-detections (i.e., upper limits based on the detection limit of the D4000-sSFR relation) along with the median $\mathrm{\Delta \Sigma_{SFR}}$ profiles for the -1.1 $< \Delta$SFR $<$ -0.5 (left panel) and $\Delta$SFR $<$ -1.1 (right panel) bins. The median profiles are the same profiles shown in Figure \ref{fig:dsfrrp}. Detections and non-detections are shown as circles and arrows, respectively, in the scatter plots. Probability density distributions of both detections (filled histogram) and non-detections (unfilled histogram) are shown along both $\mathrm{\Delta \Sigma_{SFR}}$ and $R$/$R_{\rm e}$ axes.}
    \label{fig:upperlim}
\end{figure*}

In the main body of the paper, we do not include spaxels that have D4000 $>$ 1.475 as the D4000-sSFR relation is not defined beyond this limit (Section \ref{sec:d4}). In this section, we assign upper limits to spaxels with D4000 measurements beyond the defined D4000-sSFR relation. More specifically, we assign spaxels in this regime with the sSFR value associated with the D4000 limit, which is $\log_{10}$(sSFR) $\simeq$ -11.3 yr\textsuperscript{-1}. We do this for all galaxies in our sample defined in Section~\ref{sec:metrics}. We then plot the distribution of these upper limits with that of the detections (both H$\alpha$ and D4000-based $\mathrm{\Delta \Sigma_{SFR}}$) and the median profiles for the two $\Delta$SFR bins with significant contribution of D4000 measurements: -1.1 $< \Delta$SFR $<$ -0.5 (left panel of Figure \ref{fig:upperlim}) and $\Delta$SFR $<$ -1.1 (right panel of Figure \ref{fig:upperlim}). 

For both $\Delta$SFR bins, it is expected that the majority of the upper limits are commonly distributed around a low $\mathrm{\Delta \Sigma_{SFR}}$ value ($\sim$ -1.5). In terms of distribution along the radial axes, the probabilty density distributions of detections and non-detections are strikingly similar for the -1.1 $< \Delta$SFR $<$ -0.5 bin (green), such that the peak of both distributions lie in the 0.5 -- 1.0 $R_{\rm e}$ bin. It seems fair to assume that with reliable measurements in place for these upper limits, the entire profile would be normalised towards lower $\mathrm{\Delta \Sigma_{SFR}}$. We do not expect the overall trend to change considerably. On the other hand, for the $\Delta$SFR $<$ -1.1 bin (red), non-detections are found to mainly occupy regimes around 0.5 -- 1.5 $R_{\rm e}$, whereas detections are concentrated around 1 -- 2.5 $R_{\rm e}$. This suggests that with the inclusion of upper limits, we will likely measure lower $\mathrm{\Delta \Sigma_{SFR}}$ in the centre, leading to a stronger positive gradient. Furthermore, while the number of detections vs. non-detections are comparable for the -1.1 $< \Delta$SFR $<$ -0.5 bin, the number of non-detections significantly outnumber that of the detections in the 0 - 2.0 $R_{\rm e}$ regime for the $\Delta$SFR $<$ -1.1, averaging around $\sim$2500 spaxels per bin compared to $\sim$150 spaxels for the detections. The overal normalisations of the profiles are thus upper-limits in the -1.1 $< \Delta$SFR $<$ -0.5 (green) and $\Delta$SFR $<$ -1.1 (red) bins.

\section{Bin statistics of the radial profiles}
\label{sec:app2}

This section presents information on the number of galaxies and spaxels that contribute to each radial bin for the $\mathrm{\Delta \Sigma_{SFR}}$ and Age\textsubscript{L} profiles shown in Section \ref{sec:results}. Tables \ref{tab:dssbinstat} and \ref{tab:agebinstat} show the statistics for the $\mathrm{\Delta \Sigma_{SFR}}$ and Age\textsubscript{L} profiles, respectively, where the first number of each column indicates the total number of galaxies (N\textsubscript{gal}) contributing spaxels to that bin, and the second number the total number of spaxels (N\textsubscript{spax}) in each bin.  

\begin{table*}
    \centering
    \caption{Statistics per radial bin probed by each global SF state for the $\mathrm{\Delta \Sigma_{SFR}}$ profiles, where the first number indicates the total number of galaxies in the bin, and the second the total number of spaxels.}
    \begin{tabular}{ccccc}
    \hline
        Radial bin & $\Delta$SFR $>$ 0.5 & -0.5 $<$ $\Delta$SFR $<$ 0.5 & -1.1 $<$ $\Delta$SFR $<$ -0.5 & $\Delta$SFR $<$ -1.1 \\ 
        ($R_{\rm e}$) & ($\mathrm{N_{gal}}$, $\mathrm{N_{spax}}$) & ($\mathrm{N_{gal}}$, $\mathrm{N_{spax}}$) & ($\mathrm{N_{gal}}$, $\mathrm{N_{spax}}$) & ($\mathrm{N_{gal}}$, $\mathrm{N_{spax}}$) \\
        \hline
        0.0 $<$ $R \leq$ 0.5 &  (23, 199)  &  (114, 835)  &  (25, 88)  &  $-$  \\ 
        0.5 $<$ $R \leq$ 1.0 &  (26, 586)  &  (125, 2587) &  (36, 219) &  (45, 102)  \\
        1.0 $<$ $R \leq$ 1.5 &  (26, 770)  &  (121, 3053)  &  (28, 139)  &  (50, 179)  \\ 
        1.5 $<$ $R \leq$ 2.0 &  (24, 729)  &  (91, 1745)  &  (13, 84)  &  (46, 315)  \\ 
        2.0 $<$ $R \leq$ 2.5 &  (17, 416)  &  (34, 602)  &  $-$   &  (31, 183)  \\ 
        2.5 $<$ $R \leq$ 3.0 &  (10, 104)  &  (12, 144)  &  $-$   &  (17, 77)  \\ \hline
    \end{tabular}
    \label{tab:dssbinstat}
\end{table*}

\begin{table*}
    \centering
    \caption{Statistics per radial bin probed by each global SF state for the Age\textsubscript{L} profiles, where the first number indicates the total number of galaxies in the bin, and the second the total number of spaxels.}
    \begin{tabular}{ccccc}
    \hline
        Radial bin & $\Delta$SFR $>$ 0.5 & -0.5 $<$ $\Delta$SFR $<$ 0.5 & -1.1 $<$ $\Delta$SFR $<$ -0.5 & $\Delta$SFR $<$ -1.1 \\ 
        ($R_{\rm e}$) & ($\mathrm{N_{gal}}$, $\mathrm{N_{spax}}$) & ($\mathrm{N_{gal}}$, $\mathrm{N_{spax}}$) & ($\mathrm{N_{gal}}$, $\mathrm{N_{spax}}$) & ($\mathrm{N_{gal}}$, $\mathrm{N_{spax}}$) \\
        \hline
        0.0 $<$ $R \leq$ 0.5 &  (22, 189)  &  (102, 763)  &  (25, 88)  &  $-$  \\ 
        0.5 $<$ $R \leq$ 1.0 &  (24, 560)  &  (113, 2420)  &  (36, 219)  &  (42, 97)  \\ 
        1.0 $<$ $R \leq$ 1.5 &  (24, 748)  &  (109, 2911)  &  (28, 139)  &  (50, 179)  \\ 
        1.5 $<$ $R \leq$ 2.0 &  (22, 719)  &  (81, 1677)  &  (13, 84)  &  (46, 315)  \\ 
        2.0 $<$ $R \leq$ 2.5 &  (16, 406)  &  (30, 588)  &  $-$  &  (31, 183)  \\ 
        2.5 $<$ $R \leq$ 3.0 &  $-$  &  (12, 144)  &  $-$  &  (17, 77)  \\ \hline
    \end{tabular}
    \label{tab:agebinstat}
\end{table*}

\section{Maps of measured properties}
\label{sec:app3}

\begin{figure*}
    \centering
    \begin{subfigure}{\textwidth}
        \centering
        \includegraphics[scale=0.45]{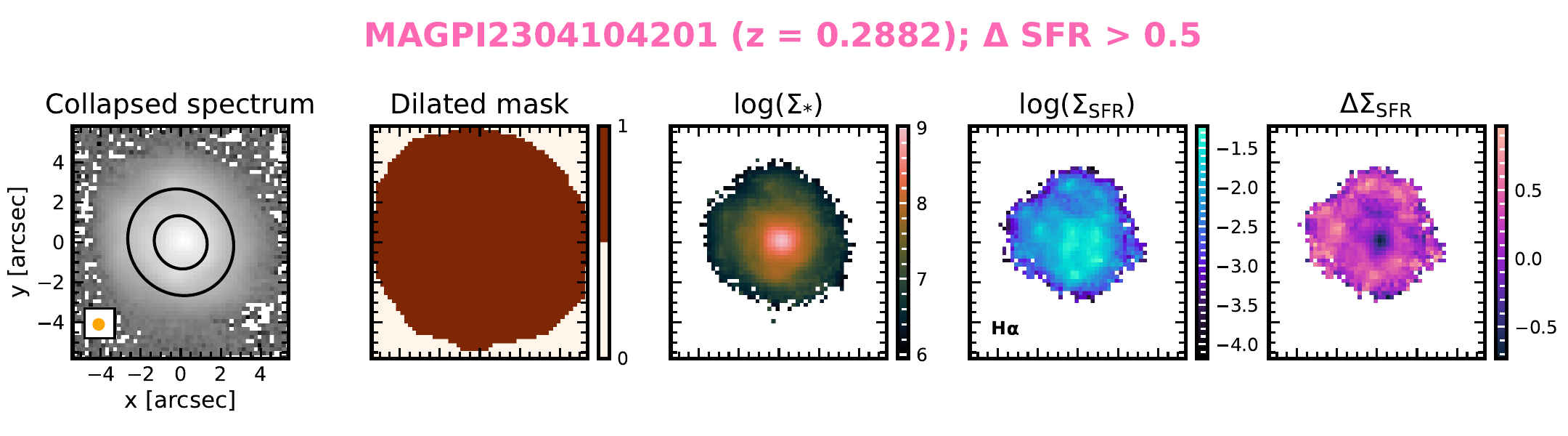}
    \end{subfigure}
    
    \begin{subfigure}{\textwidth}
        \centering
        \includegraphics[scale=0.45]{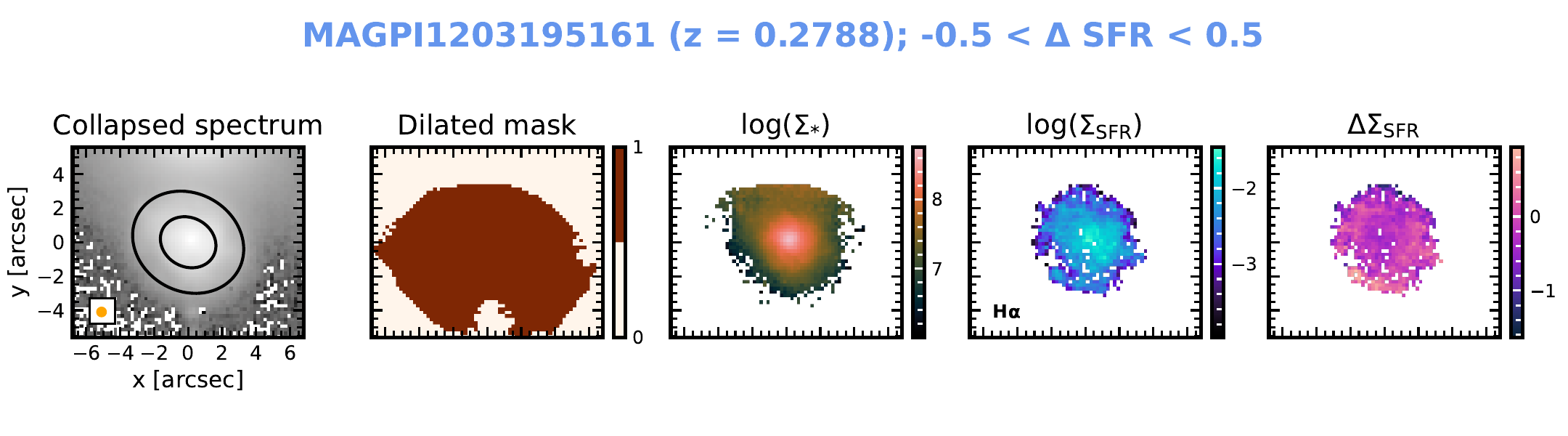}
    \end{subfigure}
    
    \begin{subfigure}{\textwidth}
        \centering
        \includegraphics[scale=0.45]{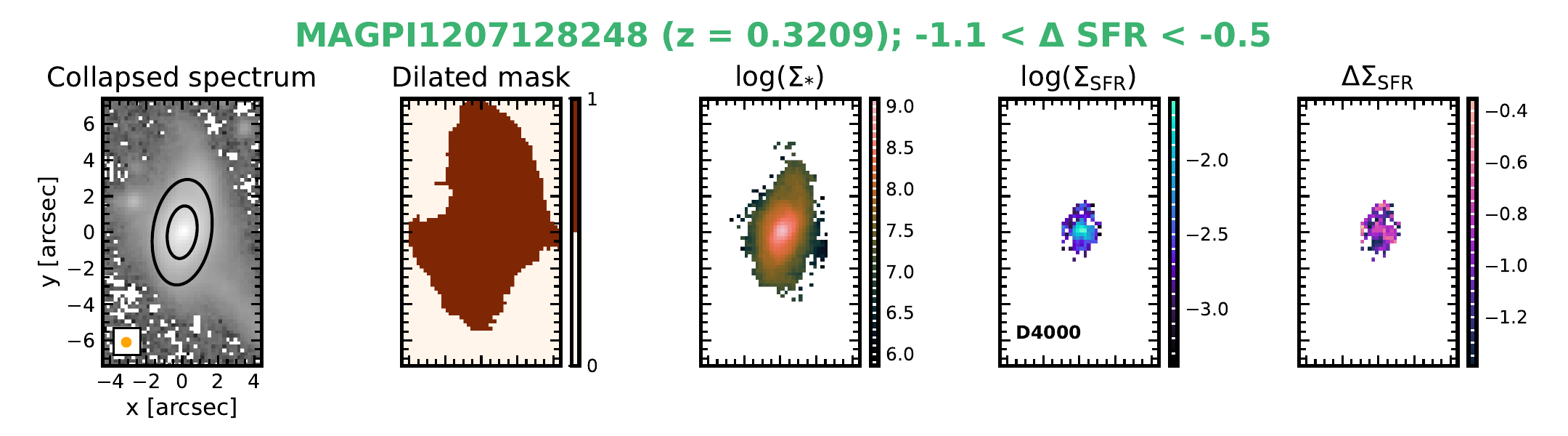}
    \end{subfigure}
    
    \begin{subfigure}{\textwidth}
        \centering
        \includegraphics[scale=0.45]{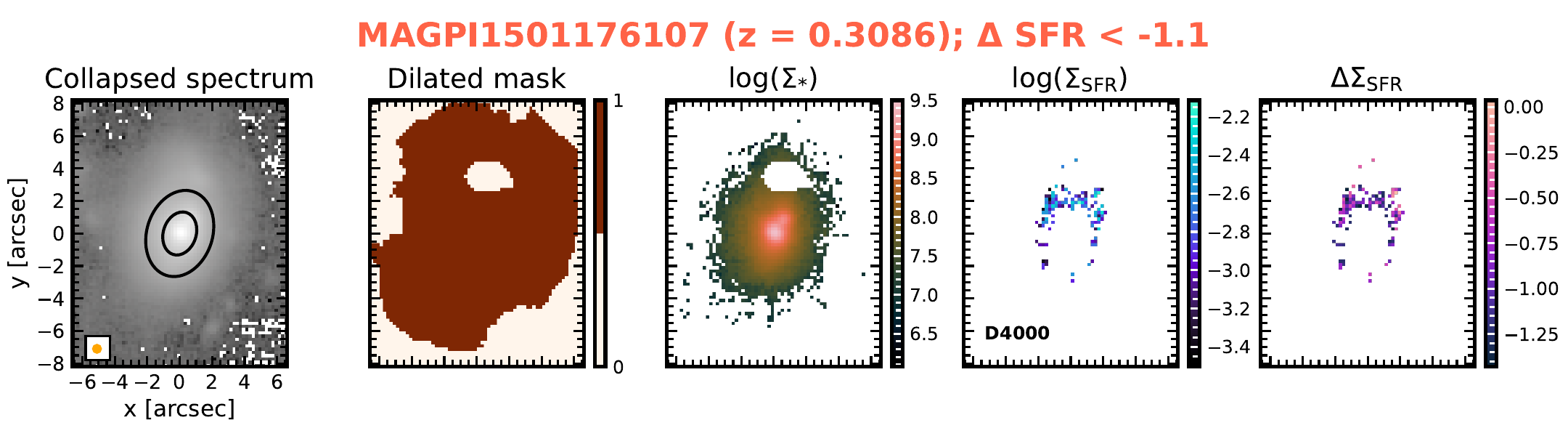}
    \end{subfigure}
    \caption{Maps of measured properties for example galaxies belonging to each global SF state, from left to right: collapsed spectrum (i.e., white light), dilated mask, stellar mass surface density ($\mathrm{\Sigma_{\star}}$ in units of $\mathrm{M_{\odot} \ kpc^{-2}}$), SFR surface density ($\mathrm{\Sigma_{SFR}}$ in units of $\mathrm{M_{\odot} \ yr^{-1} \ kpc^{-2}}$), and $\mathrm{\Delta \Sigma_{SFR}}$ (in units of dex). From top to bottom: MAGPI2304104201 ($\Delta$SFR $>$ 0.5; title coloured in pink), MAGPI1203195161 in (-0.5 $< \Delta$SFR $<$ 0.5; blue), MAGPI1207128248 in (-1.1 $< \Delta$SFR $<$ -0.5; green), and MAGPI1501176107 ($\Delta$SFR $<$ -1.1; red). The elliptical apertures of radii 1 and 2 $R_{\rm e}$, along with an orange circle denoting the PSF FWHM (average of $\sim$0.5 arcsec) measured in the $i$-band MUSE images reconstructed with the SDSS $i$-band filter transmission curve, are shown on top of the collapsed spectrum map for each row.}
    \label{fig:propmaps}
\end{figure*}

In this section, we show the maps of measured physical properties for example galaxies belonging to each global SF state in Figure \ref{fig:propmaps}. In each row of panels, we show the maps of the collapsed MUSE spectrum (i.e., white light image), dilated mask (as derived with \textsc{ProFound}, see Section \ref{sec:data}), the stellar mass surface density ($\mathrm{\Sigma_{\star}}$; see Section \ref{sec:mstar}), SFR surface density ($\mathrm{\Sigma_{SFR}}$; see Section \ref{sec:sfr}), and $\mathrm{\Delta \Sigma_{SFR}}$ (see Section \ref{sec:metrics}).

\bsp	
\label{lastpage}
\end{CJK}
\end{document}